\documentclass[12pt,a4paper]{article}
\usepackage[T1]{fontenc}
\usepackage{amsmath}
\usepackage[bitstream-charter]{mathdesign}
\usepackage{graphicx,breqn,authblk,blindtext}
\usepackage{bbm,float,longtable,mathtools,subcaption}
\usepackage{bm,algorithm,multirow,xr}
\usepackage{hyperref}
\usepackage[toc,page]{appendix}
\usepackage{theoremref,enumitem}
\usepackage[noend,compatible]{algpseudocode}
\usepackage{apacite,ntheorem,lipsum,lscape}
\usepackage[round]{natbib}

\usepackage[left=2.5cm,right=2.5cm,top=2.5cm,bottom=2.5cm]{geometry}
\DeclareMathOperator*{\argmax}{arg\,max}
\DeclareMathOperator*{\argmin}{arg\,min}
\theorembodyfont{\itshape}
\newtheorem{theorem}{Theorem}
\newtheorem{Proof}{Proof}

\newtheorem{lemma}{Lemma}

\allowdisplaybreaks

\makeatletter
\newenvironment{breakablealgorithm}
  {
   \begin{center}
     \refstepcounter{algorithm}
     \hrule height.8pt depth0pt \kern2pt
     \renewcommand{\caption}[2][\relax]{
       {\raggedright\textbf{\ALG@name~\thealgorithm} ##2\par}%
       \ifx\relax##1\relax 
         \addcontentsline{loa}{algorithm}{\protect\numberline{\thealgorithm}##2}%
       \else 
         \addcontentsline{loa}{algorithm}{\protect\numberline{\thealgorithm}##1}%
       \fi
       \kern2pt\hrule\kern2pt
     }
  }{
     \kern2pt\hrule\relax
   \end{center}
  }
\makeatother

\numberwithin{equation}{section}

\begin{document}
\date{}
\title{ MLE of Jointly Constrained Mean-Covariance of Multivariate Normal Distributions}

\author[1]{Anupam Kundu\thanks{Corresponding author, Address:155 Ireland St, College Station, TX 77840, Office: 440, Email:akundu@stat.tamu.edu}}
\author[1]{Mohsen Pourahmadi\thanks{Email:pourahm@stat.tamu.edu}}
\affil[1]{Department of Statistics, Texas A\&M University}
\maketitle

\begin{abstract}

Estimating the unconstrained mean and covariance matrix is a popular topic in statistics. However, estimation
of the parameters of  $N_p(\bm{\mu},\bm{\Sigma})$ under  joint constraints such as $\bm{\Sigma}\bm{\mu} = \bm{\mu}$ has not received much attention. It can be viewed as a multivariate counterpart of  the classical estimation problem  in the  $N(\theta,\theta^2)$ distribution.  In addition to the usual inference challenges under such non-linear constraints among the parameters (curved exponential family), one has to deal with the basic
requirements of symmetry and positive definiteness
when estimating a covariance matrix. We derive the non-linear likelihood equations for the constrained maximum likelihood estimator of $(\bm{\mu},\bm{\Sigma})$ and solve them using iterative methods.  Generally,
the  MLE of covariance matrices computed using iterative methods do not satisfy the constraints. We  propose a novel algorithm to modify
such (infeasible) estimators  or any other (reasonable) estimator. The key step is to re-align the
mean vector along the eigenvectors of the covariance matrix using the idea of regression. In using the Lagrangian function for constrained MLE \citep{aitchison1958maximum}, the  Lagrange multiplier
entangles with the parameters of interest and presents another  computational challenge. We handle this by either iterative or explicit calculation of the Lagrange multiplier. The existence and  nature of location of the constrained MLE are explored within a data-dependent convex set using recent results from random matrix theory.
 A simulation study illustrates our methodology and shows that the
modified estimators perform better than the initial estimators from the iterative methods.
\end{abstract}


\section{Introduction}
Mean and covariance estimation are of central importance in almost every area of multivariate statistics. However, estimation under joint
 constraints on the mean vector and covariance matrix of data from a  $N_p(\bm{\mu},\bm{\Sigma})$ distribution is relatively uncommon in
 multivariate statistics  \citep{bibby1979multivariate}. Our goal is to study and resolve some new challenges
 which appear when one attempts  to jointly estimate
 the mean vector and the covariance matrix of a multivariate normal distribution
  under the following two constraints:

\begin{align}
\label{ESAGCon}
\bm{\Sigma}\bm{\mu} &= \bm{\mu},  \qquad |\bm{\Sigma}| = 1.
\end{align} \par

 It is interesting to note that the first constraint forces the mean vector $\bm{\mu}$ to be
 an eigenvector of $\bm{\Sigma}$ corresponding to the
 eigenvalue one, and the second constrains the product of the remaining eigenvalues. The first constraint turns out to be more
  consequential for statistical inference due to the entanglement (nonlinearity) of the mean-covariance parameters and that
  $\bm{\mu}$ as an eigenvector is identifiable up to a constant. Nevertheless, the two together will definitely impact  the estimators and
 the shape of the contour plots of
   a multivariate normal density function as gleaned from  the spectral decomposition of the covariance matrix
 \begin{align}
 \label{SDSigma}
 \bm{ \Sigma}=\bm{PDP}^{\top}&=\sum_{i=1}^p \lambda_i \bm P_i \bm P_i^{\top}=\sum_{i=1}^{p-1} \lambda_i \bm P_i \bm P_i^{\top}+\bm{\mu}\bm{\mu}^{\top},
 \end{align} where $\bm D=\text{diag}(\lambda_1,\lambda_2,\dots,\lambda_{p-1},1)$ is the diagonal matrix of ordered  eigenvalues
   other than 1 and $\bm P=[\bm P_1,\bm P_2,\dots,\bm P_p]$ is the corresponding orthogonal matrix of eigenvectors. The second constraint is less stringent and can be
   achieved by a rescaling. Though these constraints arise in the context of directional data analysis \citep{paine2018elliptically}, they
   seem to resonate with some of the deeper  issues in the classical statistical estimation theory.

         It is well-known that constraint or functional relationship among the parameters of a distribution  can be the source of
           computational and inferential challenges. Interestingly, presence of the "quadratic" term $\bm{\mu}\bm{\mu}^{\top}$ in (\ref{SDSigma}) suggests
           similarity with the classical  inference problems for the $N(\theta,\theta^2)$ distribution
           where it is known that  the minimal sufficient
            statistic $T(X)=(\sum_{i=1}^n X_i,\sum_{i=1}^n X_i^2)$ is not complete and the UMVU estimators may not exist, see \cite{keener2011theoretical}, Chapter 5, for other interesting examples. More generally, the setup is within the multivariate curved exponential family \citep{efron1975defining} where the parameters $(\bm{\mu},\bm{\Sigma})$  satisfy the constraints in (\ref{ESAGCon}). As a potential relaxation of  the first constraint
               which is the source of most complications, and
               for the sake of demonstration, an intermediate constraint $\bm{\Sigma}\bm{b}=\bm{\mu}$ for
               some possibly known vector $\bm{b}$ is also considered, hoping that it will shed more light on the nature of the  constraints.
              We  interpret the constraints in the context of factor and
 error-in-variable models in  multivariate regression \citep{molstad2020explicit} where  the error covariance matrix
 and the regression coefficient matrix are parameterically connected.

             Though an explicit formula for the MLE of $\theta$ in $N(\theta,\theta^2)$ is given in Khan (1968), finding explicit
             formula for the MLE in our setup seems to be out of reach. A Lagrange multiplier method for computing constrained MLE and its asymptotic distribution for general
             distributions satisfying certain regularity conditions is given in \cite{aitchison1958maximum}. In this paper, focusing on multivariate
              normal distributions we incorporate the constraints in computing the MLE of the
             mean-covariance parameters, and  derive the (constrained) likelihood equations. In the absence of closed-form MLE, three iterative
              methods for computing the MLE and the Lagrange multiplier are provided and we
             study their statistical/computational  properties. Unfortunately, computing the Lagrange multiplier in our setup
             is not straightforward, perhaps due to implicit nonlinearity in the first constraint, and requires special attention.
             We compute the Lagrange multiplier using either  an iterative or explicit methods.

                          It turns out that the presumed MLEs obtained from the iterative methods invariably do not satisfy the constraint in (\ref{ESAGCon}), and in some cases the covariance estimator is neither symmetric nor positive-definite.
              It is a genuine challenge to have the MLE of the covariance matrix to satisfy (\ref{ESAGCon}), in
             addition to being  symmetric and positive definite. A novel algorithm is developed where starting with any pair of mean-covariance estimators, they
              are modified so as to satisfy the conditions in (\ref{ESAGCon}).  The key conceptual idea is to
              re-align the given  mean vector to be in the space spanned by the orthogonal eigenvectors of the given
              covariance matrix estimator.  We re-interpret this as a  regression problem with the given   mean as the response
              vector and the eigenvectors as predictors with the associated variable selection step.  The modified eigenspace
              is formed using   the Gram-Schmidt orthogonalization process starting with the given estimate of mean to
                 ensure that the estimate is an eigenvector of the estimated covariance matrix.

        The paper is organized as follows:
         Section \ref{StdEst}  provides statistical interpretation of the model  with a few  examples. Section \ref{JointMLE} describes
         three iterative
         methods of computing the  MLE and their modifications, two of the methods employ explicit calculation of the Lagrange multiplier \citep{aitchison1958maximum}. Section \ref{TheoRes} studies concavity of the Lagrangian function and provides  further theoretical justification for using the iterative methods. 
         Section \ref{LemAl} gives the details of  developing    algorithms to modify estimators satisfying both constraints, and Section \ref{SimSec} illustrates our methods through simulations. Section \ref{Conc} is the conclusion.

\section{Statistical Interpretation and Prevalence of the Constraint}
\label{StdEst}

In this section we  interpret the constraints in the context of factor and
 error-in-variable models, and then  point out that the mean-covariance of the multinomial distributions do not satisfy the constraints.\par

 Consider a factor model with a single factor of the form \citep[\S 8f.4]{rao1973linear}
 \begin{align}
\bm{X}_i &= \bm{\mu} + \bm{\mu} w_i  + \bm{\epsilon}_i
\end{align}
where $w_i \overset{iid}{\sim} N(0,1)$ and $\bm{\epsilon}_i\overset{iid}{\sim} N_p(0,\bm{\Sigma}_{\epsilon})$  are uncorrelated. Note that  the
mean vector $\bm{\mu}$  appears as the loading matrix and $w_i$ is the common factor.  The
covariance matrix of $\bm{X}_i$ is as in  (\ref{SDSigma}):
 $$\bm{\Sigma} = \bm{\Sigma}_{\epsilon} + \bm{\mu}\bm{\mu}^{\top}.$$
This factor model interpretation can also be expanded and viewed as the error-in-variable model
in the context of multivariate regression \citep{molstad2020explicit} where  the error covariance matrix
 and the regression coefficient matrix are parametrically connected. Our model  is also a special case of the
  envelop models in \cite{cook2015foundations}.

  To get a feel for the prevalence of the first constraint involving both the mean vector and the covariance matrix
   we show that multinomial distributions do not satisfy the constraints.
   In addition, we explore the role of an  "intermediate", seemingly less stringent, constraint of the form
  $\bm{\Sigma}\bm{b} = \bm{\mu}$  where $\bm{b}$ is ideally a vector independent of the parameters. However, such a $\bm{b}$ may not always exist as shown in the following example.

 Suppose $\bm{Y}\sim \text{multinomial}(n,q_1,q_2,\dots,q_p)$ where $\sum_{i=1}^p q_i = 1$. Then, $Var(Y_i) = nq_i(1-q_i)$ and $Cov(Y_i,Y_j)= -nq_iq_j$ for $i\neq j$, and  the mean and covariance matrix have the form
$$\bm{\mu}=n\bm{q},\quad\quad\bm{\Sigma}(\bm{Y}) = \text{diag}(\bm{q}) - \bm{q}\bm{q}^{\top}$$ where $\bm{q} = (q_1,q_2,\dots,q_p)^{\top}.$ The covariance matrix is positive semi-definite with one eigenvalue $0$ corresponding to the eigenvector $\mathbb{1}=(1,\dots,1)^{\top}$. We note that $\bm{\Sigma}\bm{\mu}\neq\bm{\mu}$, and there does not exist a vector $\bm{b}$ such that $\bm{\Sigma}\bm{b} = \bm{\mu}$. For example, in the one-- dimensional case $b = 1- q_1$  depends on the parameter. More
generally, the class of Dirichlet distributions is  another example of this kind which do not satisfy the constraints.

\section{Constrained Maximum Likelihood Estimation}
\label{JointMLE}

 The Lagrange multiplier method  \citep{aitchison1958maximum} is used to incorporate the constraints for
  finding the MLE of the parameters of a multivariate distribution.
 We derive the likelihood equations, present three iterative methods and study some of their computational
 and statistical properties. Curiously, the MLEs first appear to be explicit and have closed-forms, but on closer
 inspection they actually depend on the random Lagrange multipliers and hence disqualified as bona fide statistical estimators.
 This realization calls attention to estimating the Lagrange multiplier using iterative methods in conjunction with
 the MLE. Such coupling of estimation of the main and the nuisance parameters makes the task of computing the constrained MLE and study of their convergence much more challenging as shown in this section.

Let $\bm x_1,\bm x_2,\dots,\bm x_n$ be a sample of size $n$ from $N_p(\bm{\mu},\bm{\Sigma})$
 where
$\bm{\Sigma}$ is a positive-definite matrix. If  $\bm X$ is the $n\times p $ data matrix, then the log-likelihood of the multivariate normal distribution is proportional to
\begin{align}
\label{loglik}
l\left(\bm{\mu},\bm{\Sigma}|\bm X\right) &\propto   -\frac{n}{2}\log|\bm{\Sigma}| -\frac{1}{2}\sum_{i=1}^n (\bm x_i-\bm{\mu})^{\top}\bm{\Sigma}^{-1}(\bm x_i-\bm{\mu}).
\end{align}

\noindent The MLE of $(\bm{\mu},\bm{\Sigma})$ ignoring the constraints is $(\bar{\bm x},\bm S)$, the familiar sample mean and sample
 covariance matrix, which evidently do not satisfy the conditions in (\ref{ESAGCon}). However, the log-likelihood function generally is not concave under constraints on the covariance matrix and may have multiple local maxima.
 For  $n>p$ we set $\bm{A}(\bm{\mu}) = \sum_{i=1}^n (\bm{x}_i - \bm{\mu})(\bm{x}_i - \bm{\mu})^{\top}$ and note that $\bm{A}(\bm{\bar{x}}) = n\bm{S}$.

\begin{theorem}\label{thm1}
 The Lagrangian function for MLE under the intermediate constraint $\bm{\Sigma}\bm{b}=\bm{\mu}$ (expressed in terms of the inverse covariance matrix) is:
\begin{align}
\label{LagA1A2}
L\left(\bm{\mu}, \bm{\Sigma}\mid \bm{X}\right) &= l\left(\bm{\mu}, \bm{\Sigma}\mid \bm{X}\right) + \alpha_1\left(\mid\bm{\Sigma}^{-1}\mid-1\right) - \bm{\alpha}_2^{\top}\left(\bm{\Sigma}^{-1} \bm{\mu}-\bm{b}\right)
\end{align}
where $\alpha_1$ and $\bm{\alpha}_2$ are the Lagrange multipliers.
\begin{enumerate}[label=(\alph*)]
\item \label{Case1} Under the solo constraint $\mid\bm{\Sigma}\mid=1$ ($\bm{\alpha}_2 = \bm{0}$), the MLE $\widehat{\bm{\mu}}_{mle} = \bar{\bm{x}}$ is the sample mean and $\widehat{\bm{\Sigma}}_{mle} = \frac{\bm{A}(\bar{\bm{x}})}{\mid\bm{A}(\bar{\bm{x}})\mid^{1/p}}$ is a shape matrix. \\
\item \label{Case2} If $\mid\bm{\Sigma}\mid=1$ and $\bm{\Sigma}\bm{b}=\bm{\mu}$  as in  (\ref{LagA1A2}), then the constrained MLE satisfies
\begin{align}
\label{StrMLE0}
\bm{\mu} = \left(\bar{\bm{x}} - \frac{1}{n}\bm{\alpha}_2\right),\qquad & \qquad
\bm{\Sigma} = \frac{\left[\bm{A(\mu)} + 2\bm{\alpha}_2\bm{\mu}^{\top}\right]}{n+2\alpha_1}, \\  \label{ESAGConM0}
\bm{\Sigma}\bm{b} = \bm{\mu} \text{ }, \qquad & \qquad\mid\bm{\Sigma}\mid = 1.
\end{align}
\item \label{FCase} Under both constraints in (\ref{ESAGCon}), the MLE satisfies
\begin{align}
\label{StrMLE}
\bm{\mu} = \left(\bar{\bm{x}} - \frac{1}{n}\left(\bm{I} - \bm{\Sigma}\right)\bm{\alpha}_2\right),\qquad & \qquad
\bm{\Sigma} = \frac{\left[\bm{A(\mu)} + 2\bm{\alpha}_2\bm{\mu}^{\top}\right]}{n+2\alpha_1} \\
\bm{\Sigma}\bm{\mu} =\bm{\mu}\text{ }, \qquad & \qquad\mid\bm{\Sigma}\mid = 1 \nonumber
\end{align}
\end{enumerate}
\end{theorem}
The proof is provided in the Appendix \ref{PTh1}.  Unlike the closed-form solution in (a),  computing
 the MLE in (b) and (c) is more challenging and involves both $\alpha_1$ and $\bm{\alpha}_2$.
   Thus, one may resort to iterative methods for solving for the (random) Lagrange multipliers, which must
 go through all the four steps (equalities) to complete one iteration.
 To highlight the role of the intermediate constraint, we note that in Theorem \hyperref[Case2]{\ref*{thm1}.\ref*{Case2}}, every parameter can be expressed in terms of $\bm{\alpha}_2$ due to the intermediate constraint $\bm{\Sigma}\bm{b}=\bm{\mu}$, so that the iterations will be  over $\bm{\alpha}_2$ only, see Section \ref{1bal2} for details.
 By contrast,  the case  in  Theorem \hyperref[FCase]{\ref*{thm1}.\ref*{FCase}} under $\bm{\Sigma}\bm{\mu}=\bm{\mu}$  is much more challenging, at least, due to the presence of $\bm{\Sigma}$ in $\bm{\mu}$. These observations serve as strong motivations for considering the alternative
 method of  explicit calculation of the Lagrange multiplier $\bm{\alpha}_2$ in Section \ref{SAExCaLMMLE}.
  In view of Theorem \ref{thm1} (a), from here on  we focus mostly on the first constraint and deemphasize the second constraint $\mid\bm{\Sigma}\mid=1$  which is achievable through a scale change.

 \subsection{Algorithms for Computing Constrained MLE:}
 \label{PMC}
 In spite of the apparent closed forms in (\ref{StrMLE0}) and (\ref{StrMLE}), these can not be implemented or viewed as
 bona fide estimators because of their dependence on the Lagrange multipliers $\alpha_1$ and $\bm{\alpha}_2$. Here, first we
 propose a natural iterative method for computing the Lagrange multipliers leading to statistically viable estimators of
  the mean and the covariance matrix. Then, explicit calculation of the Lagrange multipliers as in \cite{aitchison1958maximum}
  and \cite{strydom2012maximum} is pursued and its role on the convergence of the iterative methods is studied.

 \subsubsection{Solving (\ref{StrMLE0}) for $\alpha_2$}
\label{1bal2}
Knowing $\bm{\alpha}_2$ in (\ref{StrMLE0}),
 determines all the other unknown quantities. To emphasize dependence on $\bm{\alpha}_2$, we set $\bm{\mu}=\bm{\mu}(\bm{\alpha}_2)$ and denote
  the numerator of $\bm{\Sigma}$ by

 $$\bm{U}(\bm{\alpha}_2)= \bm{A}\left[\bm{\mu}(\bm{\alpha}_2)\right] + 2\bm{\alpha}_2\bm{\mu}^{\top}(\bm{\alpha}_2).$$

  \noindent From the second constraint in (\ref{ESAGConM0}) it follows that $\mid\bm{U}(\bm{\alpha}_2)\mid^{1/p}= n+2\alpha_1$.
  Replacing the numerator by $\bm{U}(\bm{\alpha}_2)$ and the denominator  by $\mid\bm{U}(\bm{\alpha}_2)\mid^{1/p}$ in the right hand side of the second identity of (\ref{StrMLE0}) leads to  $$\bm{\Sigma}(\bm{\alpha}_2) = \frac{\bm{U}(\bm{\alpha}_2)}{\mid\bm{U}(\bm{\alpha}_2)\mid^{1/p}},$$
 which is a function of $\bm{\alpha}_2$.
   Substituting $\bm{\Sigma}(\bm{\alpha}_2)$ in the first expression of (\ref{ESAGConM0}),  we obtain
   \begin{align}
   \label{aleq}
   \bm{\mu}(\bm{\alpha}_2) &= \frac{\bm{\Sigma}(\bm{\alpha}_2)b}{\mid\bm{\Sigma}(\bm{\alpha}_2)\mid^{1/p}},
   \end{align}
   where further replacing  $\bm{\mu}(\bm{\alpha}_2),\bm{\Sigma}(\bm{\alpha}_2), \bm{U}(\bm{\alpha}_2)$ and $ n+2\alpha_1$ in terms of
   $\bm{\alpha}_2$ one obtains the following after some algebraic manipulation:

\begin{align}
\label{al2Sbm}
\bm{\alpha}_2 &= \frac{\mid\bm{\Sigma}(\bm{\alpha}_2)\mid^{1/p}\bar{\bm{x}} - (n-1)\bm{S}b - (1/n^2) \bm{\alpha}_2\bm{\alpha}^{\top}_2b}{2\left(\bar{\bm{x}}^{\top}b - \frac{\bm{\alpha}_2^{\top}b}{n}\right) - \frac{\mid\bm{\Sigma}(\bm{\alpha}_2)\mid^{1/p}}{n}} = f(\bm{\alpha}_2).
\end{align}

\par

 This being nonlinear in $\bm{\alpha}_2$  suggests using the iterations: $$\bm{\alpha}_2^{(k+1)} = f(\bm{\alpha}^{(k)}), k=0,1,2\dots , \text{ with } \bm{\alpha}_2^{(0)}=\bar{\bm{x}},$$
 for solving it.

   Although the intermediate constraint seems similar to (\ref{ESAGCon}), in the next section  it is demonstrated
    that the latter is much harder to work with in that one needs to iterate over the $\alpha_1$ as well.

\subsubsection{Solving (\ref{StrMLE}) for \texorpdfstring{$\alpha_1$} and \texorpdfstring{$\bm{\alpha}_2$}:}

After replacing $\bm{\mu}$ from the first identity, which involves $\bm{\Sigma}$, the second equation in (\ref{StrMLE}) reveals that $\bm{\Sigma}$ is a nonlinear function of $\bm{\alpha}_2$.  This is different from Theorem \hyperref[Case2]{\ref*{thm1}.\ref*{Case2}} in that not all  parameters can be expressed in terms of a single parameter (like $\bm{\alpha}_2$). Thus, one  may resort to iterative methods involving the four parameters $(\bm{\mu},\bm{\Sigma},\alpha_1,\bm{\alpha}_2)$ where the updates for the (k+1)-th iteration is done in the following order :
\begin{align}
\label{SMLE2}
\alpha^{(k+1)}_1 =\frac{1}{2}\left(\lvert \bm{A}\left(\bm{\mu}^{(k)}\right)+2\bm{\alpha}^{(k)}_2\bm{\mu}^{(k)\top}\rvert^{1/p} - n\right), & \qquad \bm{\Sigma}^{(k+1)} = \frac{\left[\bm{A}(\bm{\mu}^{(k)}) + 2\bm{\alpha}^{(k)}_2\bm{\mu}^{(k)\top}\right]}{n+2\alpha^{(k)}_1}\\
\bm{\alpha}^{(k+1)}_2 = \frac{1}{2}\left[\left(n+2\alpha^{(k)}_1\right)\bm{\Sigma}^{(k)} - \bm{A}\left(\bm{\mu}^{(k)}\right)\right]\bm{\mu}^{(k)},\nonumber
 & \qquad
\bm{\mu}^{(k+1)} = \bm{\Sigma}^{(k)}\left(\bar{\bm{x}} - \frac{1}{n}\left(\bm{I} - \bm{\Sigma}^{(k)}\right)\bm{\alpha}^{(k)}_2\right).
\end{align}
 Our suggested initial values are $(\bm{\Sigma}^{(0)},\bm{\alpha}_2^{(0)},\bm{\mu}^{(0)}) = (\bm{S},\bar{\bm{x}},\bar{\bm{x}})$, and for $k=0$ we compute $\alpha^{(1)}_1$ using $(\bm{\mu}^{(0)},\bm{\alpha}_2^{(0)})$ from the first equation above. But for the updates $\bm{\Sigma}^{(1)}$ and $\bm{\alpha}^{(1)}_2$, we need  the value of $\alpha_1^{(0)}$. In order to avoid the confusion, we simply choose $\alpha_1^{(0)}=\alpha_1^{(1)}$ for the first iteration, then use $(\alpha_1^{(1)},\bm{\alpha}_2^{(1)},\bm{\Sigma}^{(1)},\bm{\mu}^{(1)})$ and repeat the process.
\par

\subsubsection{ Common Challenges with Iterative Methods for Computing MLE of \texorpdfstring{$\bm{\Sigma}$}:}
An estimate of a covariance matrix from  iterative methods is usually asymmetric and not necessarily positive definite. The first
 issue is addressed by replacing the estimator  with $\frac{1}{2}\left(\bm{\Sigma}+\bm{\Sigma}^{\top}\right)$, producing an
 estimator of the form $\bm{A} + \bm{a}\bm{b}^{\top} +\bm{b}\bm{a}^{\top}$ where $\bm{a},\bm{b}\in \mathbb{R}^{p}$ and a positive definite matrix $\bm{A}$.
Ensuring positive definiteness of a matrix of this form is difficult and discussed in
the following lemma, its is presented in the Appendix.

\begin{lemma}
\label{PerMatLem} Let $\bm{a},\bm{b}\in\mathbb{R}^{p}$ and $\bm{A}$ be a positive definite matrix. Then,
\begin{enumerate}[label=(\alph*)]
\item  The non-zero eigenvalues of $(\bm{a}\bm{b}^{\top}+\bm{b}\bm{a}^{\top})$ are $\bm{a}^{\top}\bm{b} \pm \Vert \bm{a}\Vert \Vert \bm{b}\Vert$
\item \label{1NEV} The matrix $\bm{M} = \bm{A}+(\bm{a}\bm{b}^{\top}+\bm{b}\bm{a}^{\top})$ has at most one negative eigenvalue.
\end{enumerate}
\end{lemma}

To ensure positive-definiteness, Lemma \hyperref[1NEV]{\ref*{PerMatLem}.\ref*{1NEV}} suggests  replacing  the smallest eigenvalue of $\bm{M}$  by $\left(\prod_{j=1}^{p-1} \lambda_j\right)^{-1}$ where $\lambda_j$'s are the ordered eigenvalues of $\bm{M}$. This is justified by noting that according to Weyl's inequality \citep{bhatia2007perturbation}
$$\lambda_{p-1}(\bm{M})\geq \lambda_p(\bm{A})>0.$$
       \par

 In addition, there are a number of existence and convergence problems related to Theorem  \hyperref[FCase]{\ref*{thm1}.\ref*{FCase}}. These
 are dealt with partially in the next two subsections by relying on more explicit calculations of the Lagrange multipliers under the first
 constraint only.

\subsection{Explicit Calculation of the Lagrange Multiplier:}
\label{SAExCaLMMLE}

        Iterative computation  of the Lagrange multipliers along with the parameters of interest as above
        can be the source of several convergence problems. We present  a method from \cite{strydom2012maximum}
        which  computes the Lagrange multiplier  through a Taylor series expansion of the  constraint function. 
        
        Note that our  mean-covariance
        constraint can be written either as a scalar function or vector function of the parameters.
           We start with expressing the  constraint $\bm{\Sigma}\bm{\mu}=\bm{\mu}$  as the scalar function $h:\mathbb{R}^{p^2+p}\to\mathbb{R}$ of the natural parameter vector $\bm{m}$ of a multivariate normal distribution:
\begin{align}
\label{g}
h(\bm{m}) &=  \left[\bm{m}_2 - \bm{m}_1\otimes\bm{m}_1 - \text{vec}(\bm{I}_p)\right]^{\top}\left(\mathbb{1}\otimes\bm{m}_1\right)=0,
\end{align}
where $\bm{m}^{\top} = \left[\bm{\mu}^{\top},\text{vec}(\bm{\Sigma}+\bm{\mu}\bm{\mu}^{\top})^{\top}\right]^{\top}=\left[\bm{m}^{\top}_1,\bm{m}^{\top}_2\right]^{\top}$.

 Using the Taylor's expansion of $h(\bm{m})$
  around $T$, the sufficient statistics of the exponential family (the normal distribution in our case) leads to the following explicit formula for the Lagrange multiplier:
\begin{align}
\bm{\alpha}_2 &= - \left[\nabla h(\bm{m})^{\top} \nabla \bm{m}(\bm{\theta}) \nabla h(T)\right]^{-1} h(T),
\end{align}
where $\bm{\theta}$ is the canonical parameter for the multivariate normal distribution, see Appendix B.
 Substituting this in (\ref{AMTS}) leads to the identity

 \begin{align}
 \label{SANRI}
 \bm{m} &= T(\bm{X}) - \bm{V} \nabla h(\bm{m}) \frac{h(T)}{\left[\nabla h(\bm{m})^{\top} \bm{V} \nabla h(T)\right]}  \quad\text{with $\nabla \bm{m}(\bm{\theta}) = \bm{V}$}.
 \end{align}
It can be used iteratively via a "double iteration" over $T$ and $\bm{m}$, see \citet[\S 2]{strydom2012maximum}, with the initial values   chosen as the observed canonical statistics for both $T$ and $\bm{m}$, see Algorithm 2 in Appendix B.

As usual positive-definiteness and symmetry of the covariance estimate are not guaranteed. Nevertheless, its performance in terms of the Frobenius risk in the simulation studies is better than the standard MLE  procedure
 of Section \ref{PMC}. This can be attributed to the  explicit calculation of Lagrange multiplier.

\subsection{The \cite{aitchison1958maximum} Method }

    For investigating the asymptotic distribution of the MLE and its iterative computation  \citep{aitchison1958maximum}, it is common to 
     confine attention to a ball or neighbourhood of the true parameter value. More concretely, we consider the set $U_{\epsilon} = \{\bm{\theta}: \lVert \bm{\theta} - \bm{\theta}_0 \rVert <\epsilon \}$ where $\bm{\theta}_0$ is the true parameter value  for  the parameter vector  $\bm{\theta} = \left(\bm{\mu}^{\top},\text{vec}(\bm{\Sigma})^{\top}\right)^{\top}$ of
    a multivariate normal distribution. 
    
    For    
      the vector-valued constraint function
  $$h(\bm{\mu},\bm{\Sigma}) = \bm{\Sigma}\bm{\mu} - \bm{\mu},$$
   its first derivative   denoted by $\bm{H}^{1}_{\bm{\theta}}$  is the $(p+p^2)\times p$ full-rank matrix:
 \begin{align*}
 \bm{H}^{1}_{\bm{\theta}} &= \begin{bmatrix}
 \frac{\partial h}{\partial \bm{\mu}} \\
 \frac{\partial h}{\partial \bm{\Sigma}}
 \end{bmatrix} = \begin{bmatrix}
 \bm{\Sigma} - \bm{I}\\
  \bm{\mu}\otimes \bm{I}
 \end{bmatrix}.
 \end{align*}
   The notations $\bm{H}^{1}_{\hat{\bm{\theta}}}$ and $\bm{H}^{1}_{\bm{\theta}_0}$, with obvious interpretation, are used as needed next.
  The partitioned matrix $\bm{E} =  \begin{bmatrix}
\bm{B}_{\bm{\theta}_0} & -\bm{H}^{1}_{\bm{\theta}_0}\\
-\bm{H}^{1\top}_{\bm{\theta}_0} & \bm{0}
\end{bmatrix} $ is non singular \cite[Lemma 3]{aitchison1958maximum} where $\bm{B}_{\bm{\theta}_0} =\begin{pmatrix}
  \bm{\Sigma}^{-1} & \bm{0}\\
  \bm{0} & \bm{\Sigma}^{-1}\otimes \bm{\Sigma}^{-1}
  \end{pmatrix}$, and its  inverse is given by

 \begin{align*}
 \bm{E}^{-1} = \begin{bmatrix}
\bm{P}_{\bm{\theta}} & \bm{Q}_{\bm{\theta}} \\
\bm{Q}^{\top}_{\bm{\theta}}  & \bm{R}_{\bm{\theta}}
\end{bmatrix}
 \end{align*} where
 \begin{align*}
 \bm{R}_{\bm{\theta}}  = - \left(\bm{H}^{1\top}_{\bm{\theta}} \bm{B}^{-1}_{\bm{\theta}} \bm{H}^{1}_{\bm{\theta}}\right)^{-1} &= - \left[\bm{(\Sigma-I)\Sigma (\Sigma-I)} + (\bm{\mu}^{\top}\bm{\Sigma}\bm{\mu})\bm{\Sigma}\right]^{-1}\\
 \bm{Q}_{\bm{\theta}}  = - \bm{B}_{\bm{\theta}} \bm{H}^{1}_{\bm{\theta}} \bm{R} , &\quad
 \bm{P}_{\bm{\theta}}  =  \bm{B}^{-1}_{\bm{\theta}}\left[ \bm{I} - \bm{H}^{1}_{\bm{\theta}} \bm{Q}^{\top}\right].
 \end{align*}

It follows from
Lemmas 1  and  2 of \cite{aitchison1958maximum}  that, under some regularity conditions on the density and the constraint function, the solution to the equation $\frac{\partial L}{\partial \bm{\theta}}=0$ (first derivative of the Lagrangian function) exists within the set $U_{\epsilon}$ almost surely and it maximizes the likelihood function subject to the constraint $h(\bm{\theta}) = 0$.  We denote the constrained maximum likelihood estimator by  $\hat{\bm{\theta}}_n(x)$ and $\hat{\bm{\alpha}}_{2n}(x)$ for the parameters and Lagrange multiplier, respectively.
       Then, the  following joint asymptotic  normality of the estimators of the parameter vector and the Lagrange multiplier \citep{aitchison1958maximum}is useful for developing  test statistics for testing  various constraints:

\begin{align}
\begin{bmatrix}
\sqrt{n} \left(\hat{\bm{\theta}}_n - \bm{\theta}_0\right)\\
\frac{1}{\sqrt{n}} \hat{\alpha}_{2n}
\end{bmatrix} \to N\left(\bm{0},\begin{bmatrix}
\bm{P}_{\bm{\theta}}  & \bm{0}\\
\bm{0} & -\bm{R}_{\bm{\theta}}
\end{bmatrix}\right).
\end{align}
Some of the requisite regularity conditions for the above results are verified  in the Appendix \ref{PrRes} using the fact that for multivariate normal distribution all the moments exist \citep{chacon2015efficient}. The rest is verified in \cite{luo2016constrained} for sufficiently large $n$.

  Next, expressing the Taylor series expansion of the first derivative of the Lagrangian function in  matrix form, one
   arrives at the following  iterative method  \citep{aitchison1958maximum}, abbreviated as the A\&S method, for computing the MLE:
\begin{equation}
\begin{bmatrix}
\widehat{\bm{\theta}}^{(j+1)}\\
\frac{1}{n} \widehat{\bm{\alpha}}^{(j+1)}_2
\end{bmatrix} = \begin{bmatrix}
\widehat{\bm{\theta}}^{(j)}\\
 \frac{1}{n} \widehat{\bm{\alpha}}^{(j)}_2
\end{bmatrix} + \begin{bmatrix}
\bm{P}_{1\bm{\theta}} & \bm{Q}_{1\bm{\theta}}\\
\bm{Q}^{\top}_{1\bm{\theta}} & \bm{R}_{1\bm{\theta}}
\end{bmatrix} \begin{bmatrix}
\frac{1}{n} \frac{\partial l(\bm{\theta} \mid \bm{X})}{\partial \bm{\theta}}\rvert_{\bm{\theta}=\widehat{\bm{\theta}}^{(j)}} + \bm{H}_{\widehat{\bm{\theta}}^{(j)}}\frac{1}{n}\widehat{\bm{\alpha}}_2^{(j)}\\
h(\widehat{\bm{\theta}}^{(j)})
\end{bmatrix}
\end{equation} where
$\begin{bmatrix}
\bm{P}_{1\bm{\theta}} & \bm{Q}_{1\bm{\theta}}\\
\bm{Q}^{\top}_{1\bm{\theta}} & \bm{R}_{1\bm{\theta}}
\end{bmatrix} $ is the inverse of $\begin{bmatrix}
\bm{B}_{\widehat{\bm{\theta}}^{(j)}} & -\bm{H}^{1}_{\widehat{\bm{\theta}}^{(j)}}\\
-\bm{H}^{1\top}_{\widehat{\bm{\theta}}^{(j)}} & \bm{0}
\end{bmatrix} $ for $j=0$. An important point to
note here is that in the A\&S method, the coefficient matrix in the right-hand-side stays the same
through the iterations and has to invert a matrix only once.

\section{Existence  and Uniqueness of  the Constrained MLE}
\label{TheoRes}
 In this section we  study existence and  uniqueness of the constrained MLE when the search is limited to convex subsets of the parameter space.
It is based on the intuition that if the true parameter belongs to a predetermined random set  with high probability \citep{zwiernik2017maximum}, then  an iterations  restricted to this set will move closer to the true parameter.

   Recall that with the constraint $\bm{\Sigma}\bm{\mu}=\bm{\mu}$, the Lagrangian function is

\begin{align}
\label{LagM}
L\left(\bm{\mu}, \bm{\Sigma}\mid \bm{X}\right) &= l\left(\bm{\mu}, \bm{\Sigma}\mid \bm{X}\right) + \bm{\alpha}_2^{\top}\left(\bm{\Sigma} \bm{\mu}-\bm{\mu}\right)\nonumber\\
&= -\frac{n}{2}\log\mid\bm{\Sigma}\mid - \frac{1}{2}\text{Tr}\left[\bm{A}(\bm{\mu})\bm{\Sigma}^{-1}\right] + \alpha_2^{\top}\left(\bm{\Sigma} \bm{\mu}-\bm{\mu}\right)\nonumber\\
&= -\frac{n}{2}\log\mid\bm{\Sigma}\mid - \frac{n}{2}Tr\left[\bm{S}\bm{\Sigma}^{-1}\right] - \frac{n}{2}Tr\left[(\bar{\bm{x}}-\bm{\mu})(\bar{\bm{x}}-\bm{\mu})^{\top}\bm{\Sigma}^{-1}\right]+ \alpha_2^{\top}\left(\bm{\Sigma} \bm{\mu}-\bm{\mu}\right).
\end{align}
 It is not concave under the constraint on the covariance matrix and may have multiple local maxima. However, we   show that the Lagrangian function is concave
  in any direction in a predefined set of the form $\Delta_{\bm{A}} = \left\{ \bm{\Sigma}: \bm{0}\prec\bm{\Sigma}\prec{\bm{A}}\right\}$ , see \cite{zwiernik2017maximum}.
Let $\mathbb{S}^p$ denotes the set of all $p\times p$ real symmetric matrices as a subset of $\mathbb{R}^{\frac{p(p+1)}{2}}$ and $\mathbb{S}^p_{\succ 0}$ denotes the open convex cone in $\mathbb{S}^p$ of positive definite matrices. The following lemma establishes  concavity of the profiled Lagrangian function where the mean parameter is estimated by the sample mean for a fixed value of $\bm{\alpha}_2$.

\begin{lemma}
\label{DirDerLem}
 For a given value of $\bm{\alpha}_2$ and the mean vector $\bm{\mu}$ estimated by $\bar{\bm{x}}$, the
  Lagrangian function $L:\mathbb{S}^p\to\mathbb{R}$ in (\ref{LagM})  is strictly concave in $\bm{\Sigma}$ in the
  region $\Delta_{2\bm{S}}$.
\end{lemma}

  The proof  is given in the Appendix (\ref{PrRes}.\ref{PrDirDerLem}). The strict concavity of the Lagrangian function also guarantees that the covariance matrix where the Lagrangian attains its maximum is unique.
\begin{lemma}
If $\bm{\Sigma}_{max} = arg\max_{\bm{\Sigma}\in \Delta_{2\bm{S}}} L(\bar{\bm{x}},\bm{\Sigma})$, then $\bm{\Sigma}_{max}$ is unique in $\Delta_{2\bm{S}}$.
\end{lemma}

\begin{Proof}
Suppose there are two matrices $\bm{\Sigma}_1$ and $\bm{\Sigma}_{2}$ in $\Delta_{2\bm{S}}$, which maximize the Lagrangian function for a given  $\bm{\alpha}_2$. Then, for the matrix $\bm{\Sigma}(t)=(1-t)\bm{\Sigma}_1+t\bm{\Sigma}_2$, $t\in [0,1]$, we have

\begin{align*}
L(\bar{\bm{x}},\bm{\Sigma}(t))&\geq (1-t)L(\bar{\bm{x}},\bm{\Sigma}_1)+tL(\bar{\bm{x}},\bm{\Sigma}_2) = L(\bar{\bm{x}},\bm{\Sigma}_1)
\end{align*}
 so that $\bm{\Sigma}(t)$'s also maximizes the Lagrangian function. Therefore there is a direction in which the Lagrangian is not strictly concave contradicting lemma \ref{DirDerLem}. So if the maximizer exists within $\Delta_{2\bm{S}}$, it is unique.
\end{Proof}

 To analyze the probability that $\Delta_{2\bm{S}}$ contains the true covariance matrix, we rely
          on the known fact that \citep[Theorem 3.4.1]{bibby1979multivariate} a sample covariance matrix $\bm{S}$ based on a random sample of $n\geq p$ observations from $N_p(\bm{\mu},\bm{\Sigma})$, follows a Wishart distribution i.e. $n\bm{S}\sim \pi_{W}(n-1,\Sigma)$ and also $\bm{W}_{n-1}=n\bm{\Sigma}^{-1/2}\bm{S}\bm{\Sigma}^{-1/2}\sim \pi_{W}(n-1,\bm{I}_p)$. Then, the probability that $\bm{\Sigma}\in \Delta_{2\bm{S}}$ is expressed as follows:

  \begin{align*}
  P\left[\bm{\Sigma}\in \Delta_{2\bm{S}}\right] &= P\left[2\bm{S}-\bm{\Sigma}\succ \bm{0}\right]= P\left[2\bm{\Sigma}^{-1/2}\bm{S}\bm{\Sigma}^{-1/2}-\bm{I}_p\succ \bm{0}\right]\\
  &= P\left[\frac{2}{n}W_{n-1}\succ\bm{I}_p\right]=P\left[\bm{W}_{n-1}\succ \frac{n}{2}\bm{I}_p\right]\\
  &=P\left[\lambda_p(\bm{W}_{n-1})> \frac{n}{2}\right].
  \end{align*}
                  Interestingly, the probability that the true parameter $\bm{\Sigma}$ lies within the set $\Delta_{2\bm{S}}$ is independent of $\bm{\Sigma}$ and is equal to the probability that $\lambda_p(\bm{W}_{n-1})>\frac{n}{2}$ where $\bm{W}_{n-1}\sim \pi_W(n-1,\bm{I}_p)$. It is known that \citep{zwiernik2017maximum} this probability gets closer to 1 as $n, p\to \infty$,$n/p \to \gamma^*<6+4\sqrt{2}$. Thus, for big enough dataset we expect an iterative algorithm, when restricted to this random set, will eventually converge to the constrained MLE.

\section{ An Algorithm for Enforcing the Constraints}
\label{LemAl}

   Most estimators presented so far do not necessarily satisfy the constraints. In this section, starting with any reasonable estimators for $(\bm{\mu}, \bm{\Sigma})$ (like those in Sections \ref{JointMLE}), we present an algorithm
    for modifying them so as to satisfy both constraints in (\ref{ESAGCon}).  The notation $M_0=(\tilde{\bm{\mu}},\tilde{\bm{\Sigma}})$ is used from here on to denote any such pre-estimate of $(\bm{\mu},\bm{\Sigma})$ and $M_i, i=1,2,3$ for its gradual modifications.

\subsection{  Scale Modifications of the Mean and Covariance Matrix}
The modification process starts by the  task of modifying the given covariance matrix estimator
  to accommodate the mean vector estimate. For $p=3$, a slightly different reparameterization of the
covariance matrix is developed in \cite{paine2018elliptically}.

\begin{lemma}
\label{Algo}
Given $\tilde{\bm{\mu}}\in\mathbb{R}^p$ and $\tilde{\bm{\Sigma}}$ any $p\times p$ positive-definite covariance matrix with
 the spectral decomposition $\bm{PDP}^{\top}$ as in (\ref{SDSigma}). Set $\bm P^*_p = \frac{\tilde{\bm{\mu}}}{||\tilde{\bm{\mu}}||}$ and apply the
Gram - Schmidt orthonormalization process to the set of vectors $\{\bm P^*_p,\bm P_{p-1},\dots,\bm P_2,\bm P_1\}$
to obtain $\{\bm P^*_p,\dots,\bm P^*_2,\bm P^*_1\}$.  Then, the modified covariance matrix
\begin{align}
\label{AlgoE}
\widehat{\bm{\Sigma}}^* = \sum_{j=1}^{p-1} \frac{\lambda_j}{\lambda_{pr}}\bm P^*_j\bm P^{*\top}_j + \bm P^*_p \bm P_p^{*\top}\qquad \text{ where } \lambda_{pr} = \left(\prod_{k=1}^{p-1} \lambda_k\right)^{\frac{1}{p-1}}
\end{align}
satisfies the conditions in (\ref{ESAGCon}).
\end{lemma}

We denote this estimator by $M_1$. In Lemma \ref{Algo}, $\tilde{\bm{\mu}}$ is effectively forced to become an eigenvector corresponding to the eigenvalue 1 of
 a modified  covariance matrix estimator, i.e.
$\widehat{\bm{\Sigma}}^*\tilde{\bm{\mu}}=\tilde{\bm{\mu}}$.
 It turns out that estimators obtained by this simple-minded modification, and inspired by basic linear algebra do not perform well.  This is somewhat expected as only the  covariance estimator is modified and the mean vector
            is left intact.

In view of the  simultaneous constrains on the mean vector and the covariance matrix, their joint modification seems a  natural idea to consider. Next, the  mean vector is forced in  the direction (span) of the eigenvectors of the  covariance estimator. This  is implemented by entertaining regression-like models for the given mean  vector with the
 eigenvectors serving as covariates. First, we consider simple linear regressions by choosing a single eigenvector and estimating the corresponding regression coefficient $c$ i.e. $\tilde{\bm{\mu}}=c\bm P_i$ for some eigenvector $P_i$.

\begin{lemma}
\label{AlgoM}
Given $\tilde{\bm{\mu}}\in\mathbb{R}^p$ and $\tilde{\bm{\Sigma}}$  a $p\times p$ positive-definite covariance matrix with spectral decomposition $\bm{PDP}^{\top}$. Define

\begin{align}
\label{AlgoMEx}
c_{0i}=\argmin_{c\in \mathbb{R}} \left\Vert \tilde{\bm{\mu}} - c\bm P_i\right\Vert ^2 = \frac{\langle \bm P_i,\tilde{\bm{\mu}}\rangle}{||\bm P_i||^2}\qquad &\text{ and }\qquad i_0=\argmin_{i} \left(1-\frac{\lambda_i}{c^2_{0i}}\right)^2.
\end{align}
Then, the modified mean-covariance estimators
\begin{align}
\label{L2Est}
\widehat{\bm{\mu}}^* = c_{0i_0} \bm P_{i_0} &\text{,}\quad
\widehat{\bm{\Sigma}}^* = \sum_{\substack{j\neq i_0 \\ j=1}}^{p} \frac{\lambda_j}{\lambda_{pr}}\bm P_i\bm P^{\top}_i + \widehat{\bm{\mu}}^* \widehat{\bm{\mu}}^{*\top} \text{ where } \lambda_{pr} = \left(\prod_{\substack{j\neq i_0 \\ j=1}}\lambda_j\right)^{\frac{1}{p-1}}
\end{align}
satisfy (\ref{ESAGCon}).
\end{lemma}

We refer to the  estimator $(\widehat{\bm{\mu}}^*,\widehat{\bm{\Sigma}}^*)$ as $M_2$ in the sequel. The intuition behind  the method for selecting $i_0$ is that from $\widehat{\bm{\mu}}^*\widehat{\bm{\mu}}^*=c^2_{0i}\bm P_i\bm P_i^{\top}$ it is desirable to have the eigenvalue corresponding
 to $\widehat{\bm{\mu}}^*$ to be as close as possible to one of the $\lambda_i$'s. Thus,  it is  reasonable that $\lambda_i/c^2_{0i}$ should
  be as close to 1 as possible. More details about such selection can be found  in Appendix (\ref{L2AlgoMEx}).\par

Modifying the initial estimator jointly using (\ref{L2Est}) we obtain $(\widehat{\bm{\mu}}^*,\widehat{\bm{\Sigma}}^*)$. Since the covariance matrix is not modified too much it is likely that the mean will suffer too much while the estimator of the covariance will not. In light of this intuition we need to have a balance for joint estimation while satisfying the constraint.

\subsection{The Modification Algorithm: Multiple Regression}

   In this section we consider a full-fledged multiple linear modeling of $\tilde{\bm{\mu}}$ on $\bm P_i$'s. It  amounts to a generalization of  Lemma \ref{AlgoM}     and  involves variable selection in the context of multiple regression. 
The    details  are  organized in the following Algorithm \ref{AlgoM2}, where the task is to divide the eigenvectors (regressors) into two groups. We rely  on the maximum distance between the consecutive
     terms of ordered absolute values of the regression coefficients in the saturated model. A viable alternative for this is the 2-means clustering
     algorithm  applied to absolute values of the entries of the vector $c$ of regression coefficients.  The estimator from this algorithm is denoted by $M_3$.

\begin{breakablealgorithm}
\caption{ Modifying an Estimator to Satisfy (\ref{ESAGCon})}
\label{AlgoM2}
\begin{algorithmic}[1]
\State Start with a given $(\tilde{\bm{\mu}},\tilde{\bm{\Sigma}})$ and its spectral decomposition as in (\ref{SDSigma})
\State \textbf{Variable (Basis) Selection:}
Write $\tilde{\bm{\mu}}=\sum_{j=1}^p c_i\bm P_i = \bm{Pc}$ where $c=(c_1,c_2,\dots,c_p)$.
\begin{itemize}
\item Simple Clustering : Viewing the $c_i$'s as weights, select those $P_i$'s which has largest absolute weight by ordering absolute values of $c_i$'s and find out the biggest gap. Let the index set of the group with higher absolute value of $c_i$ be
$\mathcal{S}=\{i_1,i_2,\dots,i_{j_0}\}$ where $j_0$ is its cardinality.

\textbf{OR}

\item Cluster $c_i$'s by applying K-means clustering with $K=2$ \citep{hartigan1979algorithm} on absolute values of $c_i$'s.
\end{itemize}

\State\textbf{Regress $\tilde{\bm{\mu}}$ on the span of columns of $\bm P_{j_0}=[\bm P_{i_1},\bm P_{i_2},\dots,\bm P_{i_{j_0}}]$:}
\begin{align}
\label{muEs}
\widehat{\bm{\beta}}=\argmin_{\beta} \left\Vert\tilde{\bm{\mu}}-\bm P_{j_0}\bm{\beta}\right\Vert^2,\quad\widehat{\bm{\mu}}^*&=\bm P_{j_0}\widehat{\bm{\beta}}
\end{align}

\State \textbf{Orthogonalization to accommodate $\widehat{\bm{\mu}}^*$:}  Let $g=\argmax\{\mid\widehat{\beta}_{k^*}\mid: k^*=i_1,i_2,\dots,i_{j_0}\}$. Apply the Gram-Schmidt process  on $\left\{\bm P_{i_1},\dots,\bm P_{i_{g-1}},\widehat{\bm{\mu}}^*,\bm P_{i_{g+1}},\dots,\bm P_{i_{j_0}}\right\}$ to obtain $\left\{\widehat{\bm{\mu}}^*,\bm b_1,\bm b_2,\dots,\bm b_{j_0-1}\right\}$ with $\widehat{\bm{\mu}}^*$ as the starting vector.
\State
Set, $$\widehat{\bm{\Sigma}}^*= \widehat{\bm{\mu}}^* \widehat{\bm{\mu}}^{*\top} +\sum_{\substack{j\neq \{i_1,\dots,i_{j_0}\} \\ j=1}}^p \lambda_j \bm P_j\bm P^{\top}_j + \sum_{k=1}^{j_0-1} \lambda^'_{i_k} \bm b_k \bm b_k^{\top}$$
estimate $\lambda_{i_k}^'$ by

\begin{align}
\label{LEs}
\widehat{\lambda}_{i_k}=\bm b^{\top}_k\bm{\Sigma} \bm b_k , \qquad  k=1,2,\dots, j_0-1.
\end{align}
(The proof of this step is presented in Appendix \ref{EVEstPr}).

\State Let $\lambda_{pr}=\left(\prod_{\substack{j\not\in \mathcal{S} \\ j = 1}}^p \lambda_j.\prod_{k=1}^{j_0}\widehat{\lambda}_{i_k}\right)^{\frac{1}{p-1}}$. The modified estimator $(\widehat{\bm{\mu}}^*,\widehat{\bm{\Sigma}}^*)$ is given by

\begin{align}
\label{AEst}
\widehat{\bm{\mu}}^*&=\bm P_{j_0}\widehat{\bm{\beta}}\nonumber\\
\widehat{\bm{\Sigma}}^*&=\sum_{\substack{j\neq \{i_1,\dots,i_{j_0+1}\} \\ j=1}}^p \frac{\lambda_j}{\lambda_{pr}} \bm P_j\bm P^{\top}_j + \sum_{k=1}^{j_0} \frac{\widehat{\lambda}_{i_k}}{\lambda_{pr}} b_k b_k^{\top}+ \widehat{\bm{\mu}}^* \widehat{\bm{\mu}}^{*\top}.
\end{align}
\end{algorithmic}
\end{breakablealgorithm}

\section{Simulation Experiments}

\label{SimSec}

 Through several simulation experiments, we assess the performance of the following three iterative methods and our modified estimators:  1. Standard MLE, 2. Standard MLE with explicit calculation of Lagrange multiplier (denoted by S\&C), 3.  The \cite{aitchison1958maximum} iterative method (denoted by A\&S).

\subsection{ The Simulation Set up:}
  We have taken sample size and dimension to be $(n,p) = (50,5), (50,25), (100,10), (300,30)$. Risks are approximated by averaging the losses for 100 independent replications in each of the four combinations of $(n,p)$. In all cases the data generation mechanism and  the risk function are kept the same, we have used Frobenius loss as our default loss function and calculated Stein's loss in some specific cases.

 For the parameters of the Gaussian distributions used for data generation we  take the entries of the mean vector $\bm{\mu}$  to be values of independent  standard Gaussian variables. For the covariance matrix, we start with
   $\bm{\Psi}=\bm{LL}^{\top}$ where $\bm{L}$ is a lower triangular matrix with the diagonal entries generated from $N(5,1)$ and standard normal
   for the off-diagonal entries.  The larger
 diagonal entries of  $\bm L$ ensure positive-definiteness of $\bm{\Sigma}$. Since such $(\bm{\mu},\bm{\Psi})$ do not necessarily satisfy conditions (\ref{ESAGCon}), the
    covariance matrix is modified first by applying (\ref{AlgoE}) to $(\bm{\mu},\bm{\Psi})$.

The performance of the estimators is assessed using the scaled $L_2$ risk \citep[\S 3.1]{ledoit2004well}:

$$R(\bm{\mu},\widehat{\bm{\mu}}^*)=\mathbb{E}\left[\frac{1}{p}\left\Vert\widehat{\bm{\mu}}^*-\bm{\mu}\right\Vert^2_{\mathcal{F}}\right]\qquad,\qquad R(\bm{\Sigma},\widehat{\bm{\Sigma}}^*)=\mathbb{E}\left[\frac{1}{p}\left\Vert\widehat{\bm{\Sigma}}^*-\bm{\Sigma}\right\Vert^2_{\mathcal{F}}\right],$$
     where
   $\widehat{\bm{\mu}}^*$ and $\widehat{\bm{\Sigma}}^*$ are the final modified estimators described in Section \ref{LemAl}.

 \subsection{ Simulation Results from the Three Iterative Methods:}

\subsubsection{ The Standard MLE:}
\label{SCSMLE}

   The iterative  method for computing the maximum likelihood estimator
    described in Section \ref{PMC}
   does not always converge.

   Since convergence of the  four sets of
    parameters simultaneously is unlikely, the convergence criterion  used here is
     to stop iterations  if at least two of the parameters converge.
     In most  cases the iterations for $\bm{\mu}$ and $\alpha_1$ converge, but the rate of decrease of Frobenius risk for estimating $\bm{\Sigma}$ is slow in successive iteration. In the simulations  we have taken the maximum number of iterations to be 1000. When the convergence does not happen after 1000 iterations,  we take the  output at the 1000-th iteration  as the estimator and pass it through the Algorithm 1 for $M_3$ to arrive at the final estimator.   This  method  referred to as the standard MLE (SMLE), involves iterative updating of the Lagrange multipliers.
     In contrast, the next two iterative methods involve exact calculation of the Lagrange multiplier.

 \subsubsection{The S\&C Method:}

 The S\&C method is described in Section \ref{SAExCaLMMLE}. It  does not guarantee the positive definiteness of the estimate of the covariance matrix. Thus,  we  only take the cases where the estimate is  positive definite for the  risk calculation, otherwise the corresponding simulation run is ignored (see table \ref{tab:PDFreq-SC})

\begin{table}[H]
\centering
\caption{No of Times the Estimate is Positive Definite}
\label{tab:PDFreq-SC}
\begin{tabular}{|c|c|c|}
\hline
\textbf{n} & \textbf{p} & \textbf{\begin{tabular}[c]{@{}c@{}}No of cases with\\ positive definite\\ covariance estimate\end{tabular}} \\ \hline
50  & 5  & 90  \\ \hline
50  & 25 & 100 \\ \hline
100 & 10 & 99  \\ \hline
300 & 30 & 100 \\ \hline
\end{tabular}
\end{table}

  Moreover, the  method does not guarantee  exact satisfaction of the constraints, so we apply the  Algorithm \ref{AlgoM2}
   to the estimates using $M_3$ with two types of clustering, they produce similar results with K-Means clustering performing slightly better.

  Since  convergence is a  recurring issue, we have taken the maximum number of iterations in both the loops
    of the "double iteration" to be 100, and the value of $\epsilon$ to be $0.1$. From the Table \ref{CombTab} we can see that the S\&C method  is losing very little while achieving the satisfaction of  the joint constraint (\ref{ESAGCon}).

\subsubsection{The A\&S Method:}

The iterative method for calculation of the constrained MLE described in Section \ref{TheoRes} operates inside a closed ball of radius $\delta = \Vert \bm{\theta}^{(0)} \Vert$ around the true parameter.
 Hence choosing a good initial value for the iterations to run is essential  and here we  chose  $\bm{\theta}^{(0)} = \left(\bar{x}^{\top},\text{vec}(S)^{\top}\right)^{\top}$ .

Suppose in the $i$-th stage we have the value of the parameter vector to be $\bm{\theta}^{(i)}_0 = \bm{a}$ and in the $(i+1)$-th step it moves to a point $\bm{\theta}^{(i+1)}_0 = \bm{b}$  outside the ball. Then, we find  the point $\bm{c} = (1-t) \bm{\theta}^{(0)} + t\bm{b}$ with  $t = \frac{\delta}{\left\Vert \bm{\theta}^{(0)}-\bm{b} \right\Vert}$ resides on the ball, and continue the iteration with
 the new point $\bm{c}$ instead of $\bm{b}$.  This can be seen from the picture.

  \begin{figure}[H]
  \centering
   \includegraphics[width=0.6\textwidth]{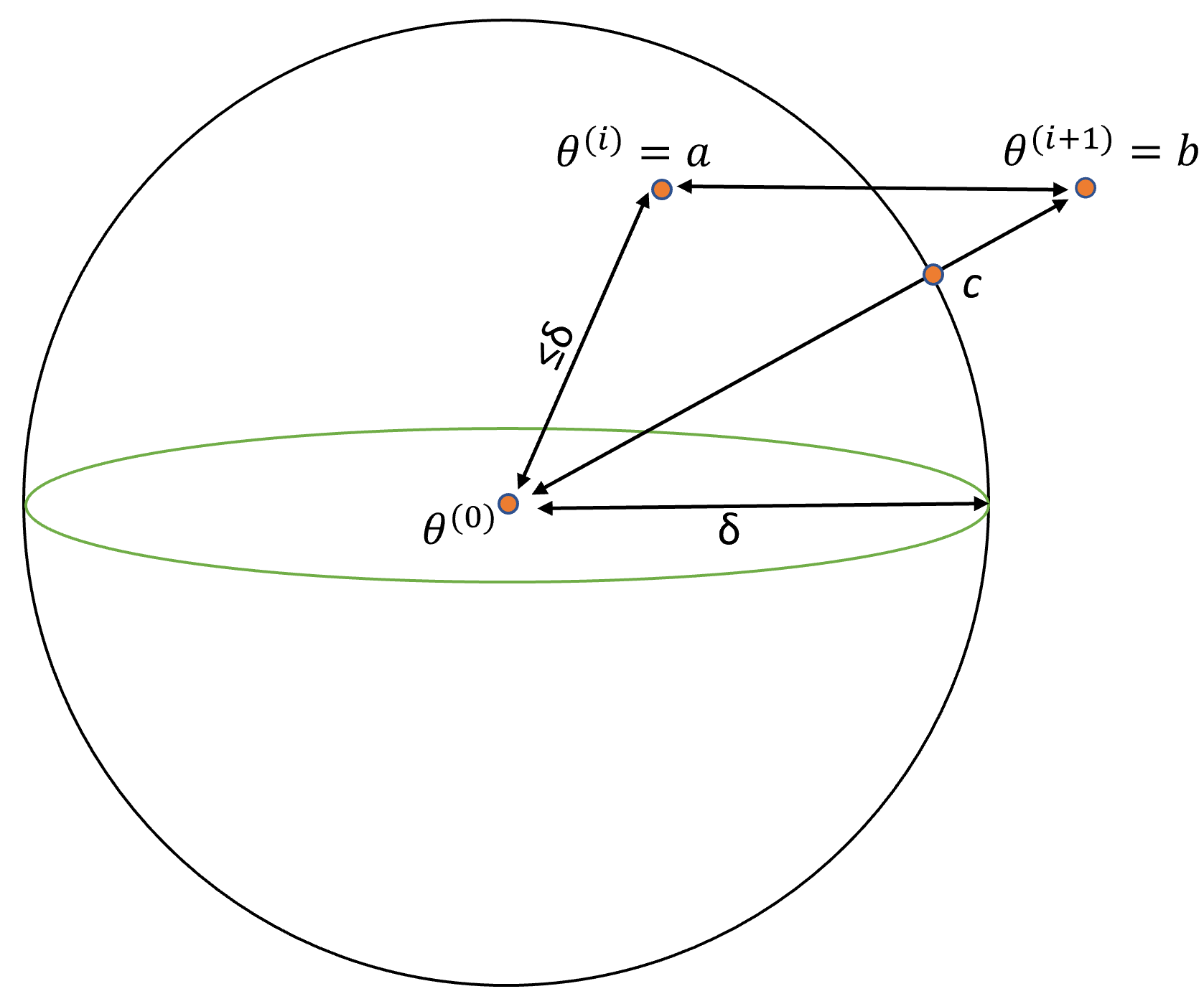}
   \caption{This pictorial representation shows how we update when the iteration goes outside the ball in \cite{aitchison1958maximum} method}
  \end{figure}

  In each iteration we symmetrize the update for the covariance matrix.
                        We  take only those simulation runs where iterations produce a positive definite output. The interesting part is that the positive definiteness is achieved after symmetrization in almost all.
                         The performance is close to the S\&C method as can be seen from table \ref{CombTab}.

\begin{table}[H]
\centering
\caption{Risks for the three iterative methods of finding
constrained MLE, modified by Algorithm 1 (M3) with K-Means.  SMLE: standard MLE; S\&C is the method of \cite{strydom2012maximum},
 and A\&S denotes the method of \cite{aitchison1958maximum}. }
\label{CombTab}
\begin{tabular}{|c|c|c|c|c|c|c|}
\hline
\textbf{} & \textbf{} & \textbf{} & \multicolumn{2}{c|}{\textbf{Mean}} & \multicolumn{2}{c|}{\textbf{Sigma - Frobenius}} \\ \hline
\textbf{Method} & n & p & MLE & \textit{M3} & MLE & \textit{M3} \\ \hline
SMLE & 50 & 5 & 0.5313 & 0.5463 & 1.2632 & 0.4212 \\ \hline
\textbf{S\&C} & \textbf{} & \textbf{} & 0.4817 & 0.5206 & 0.3553 & 0.3057 \\ \hline
A\&S &  &  & 0.4389 & 0.6296 & 1.2246 & 0.6097 \\ \hline
SMLE & 50 & 25 & 0.7998 & 0.8 & 6.1632 & 1.5567 \\ \hline
S\&C &  &  & 0.7913 & 0.8158 & 1.073 & 1.5678 \\ \hline
\textbf{A\&S} & \textbf{} & \textbf{} & 0.1691 & 0.8643 & 2.4889 & 2.3021 \\ \hline
SMLE & 100 & 10 & 0.6709 & 0.673 & 2.3789 & 0.4617 \\ \hline
S\&C &  &  & 0.6468 & 0.6849 & 0.3797 & 0.3963 \\ \hline
\textbf{A\&S} & \textbf{} & \textbf{} & 0.2814 & 0.6797 & 1.5451 & 0.8660 \\ \hline
SMLE & 300 & 30 & 0.8071 & 0.8072 & 4.9507 & 0.5507 \\ \hline
S\&C &  &  & 0.8014 & 0.8238 & 0.5117 & 0.5477 \\ \hline
\textbf{A\&S} & \textbf{} & \textbf{} & 0.1575 & 0.9639 & 2.1536 & 0.8122 \\ \hline
\end{tabular}
\end{table}

\subsection{An Example: Estimates of the Historic Position of
Earth’s Magnetic Pole}

 The dataset collected by \cite{schmidt1976non} contains the site mean direction estimates of the Earth's historic magnetic pole from 33 different sites in Tasmania. The longitude and latitudes from the data set is transformed to $X_1,X_2,\dots, X_{33}$ on a three dimensional unit sphere \citep{preston2017analysis}. The angualr gaussian distribution family is the marginal directional component of a multivariate normal distribution with ESAG distribution as a subfamily. \cite{paine2018elliptically} provided strong evidence in favor of ESAG distribution which satisfy the constraint over isotropic angular gaussian distribution while analysing this dataset. This inspire us to make normality assumption under the constraint similar to ESAG distribution disregaring the spherical nature of the tansformed dataset. The constrained maximum likelihood estimate calculated using the numerical method by \cite{aitchison1958maximum} with 1000 iterations is: 
 
 $$\bm{\mu} =\begin{bmatrix}
  -0.593 & 0.167 & 0.787 
 \end{bmatrix}, \quad \bm{\Sigma} = \begin{bmatrix}
 0.670  & 0.235 & -0.299\\
 0.235  & 2.333 & -0.106\\
 -0.299 & -0.106 & 0.797
 \end{bmatrix} $$

 \noindent which is comparable to the maximum likelihood estimate calculated using eeliptically symmetric angular Gaussian distribution with a specific parametrization in three dimension \citep{paine2018elliptically}. One main advantage is that our calculation is not restricted to three dimension.

\section{Conclusions}
\label{Conc}
 We address construction of a joint estimator for the mean-covariance of a normal distribution under the constraints (\ref{ESAGCon}).
  Three iterative methods are presented where the end results do not necessarily satisfy the constrains or the basic requirements
  of being a covariance matrix. Our novel  algorithm
   modifies any joint estimator of the mean-covariance to satisfy the constraints.
                           Comparison of the three methods for finding constrained maximum likelihood estimator shows
                           an advantage for explicit computation of the Lagrange multiplier, possibly because the corresponding
                            iterative methods  are variants of the Newton- Raphson  algorithm.
As for future research directions, it is of interest to find the maximum likelihood estimators under the constraint $\bm{\Sigma}\bm{\mu} = \bm{\mu}$ for broader class of distributions such as the elliptically contoured distributions, and  testing the validity of the hypothesized constraints.

\vspace*{1cm}

\noindent\textbf{Funding:} This research did not receive any specific grant from funding agencies in the public, commercial, or
not-for-profit sectors.

\noindent\textbf{Declarations of interest:} none

\bibliographystyle{apalike}

\bibliography{../../../Reffiles/Ref}

\newpage
\appendix
\appendixpage

\section{Proofs of Results:}
\label{PrRes}
\begin{enumerate}

\item \label{PTh1}\textbf{Proof of Theorem \ref{thm1}:}
\begin{enumerate}[label = (\alph*)]
\item By differentiating the Lagrangian in \hyperref[Case1]{\ref*{thm1}.\ref*{Case1}}
\begin{align}
L(\bm{X};\bm{\mu},\bm{\Sigma}) &= l(\bm{X};\bm{\mu},\bm{\Sigma}) + \alpha_1(\mid\bm{\Sigma}^{-1}\mid-1)\nonumber\\
& = c + \frac{n}{2}\log\mid\bm{\Sigma}^{-1}\mid - \frac{1}{2}Tr\left[\bm{A}(\bm{\mu})\bm{\Sigma}^{-1}\right] + \alpha_1\left(\mid\bm{\Sigma}^{-1}\mid - 1\right)
\end{align}

  with respect to $\bm{\mu}$ and setting to zero leads to $\widehat{\bm{\mu}}_{mle} = \bar{\bm{x}}$. Differentiation with respect to $\alpha_1$ gives us the condition $ \mid\bm{\Sigma}^{-1}\mid = 1$. Using this and setting derivative with respect to $\bm{\Sigma}^{-1}$ to 0, leads to

\begin{align}
\frac{\partial L(\bm{X};\bm{\mu},\bm{\Sigma}) }{\partial \bm{\Sigma}^{-1}} &= \frac{1}{2} \left[n\bm{\Sigma} - \bm{A}(\bm{\mu}) + 2\alpha_1 \mid\bm{\Sigma}^{-1}\mid \bm{\Sigma}\right]\nonumber\\
&=  \frac{1}{2} \left[n\bm{\Sigma} - \bm{A}(\bm{\mu}) + 2\alpha_1 \bm{\Sigma}\right]\nonumber\\
\widehat{\bm{\Sigma}} &= \frac{\bm{A}(\bm{\mu})}{\mid\bm{A}(\bm{\mu})\mid^{1/p}}
\end{align}

Now substituting for the MLE of $\bm{\mu}$ we obtain $\widehat{\bm{\Sigma}}_{mle} =  \frac{\bm{A}(\bar{\bm{x}})}{\mid\bm{A}(\bar{\bm{x}})\mid^{1/p}}. $

\item The Lagrangian in \hyperref[Case2]{\ref*{thm1}.\ref*{Case2}} is:
\begin{align}
L(\bm{X};\bm{\mu},\bm{\Sigma}) &=c + \frac{n}{2}\log\mid\bm{\Sigma}^{-1}\mid - \frac{1}{2}Tr\left[\bm{A}(\bm{\mu})\bm{\Sigma}^{-1}\right] + \alpha_1(\mid\bm{\Sigma}^{-1}\mid-1) - \bm{\alpha}^{\top}_2\left(\bm{\Sigma}^{-1}\bm{\mu} - b\right)
\end{align}
  Rewriting the condition as $\bm{\Sigma}^{-1}\bm{\mu} =\bm{b}$ and $\lvert\bm{\Sigma}^{-1}\rvert = 1$  in the Lagrangian  is necessary for taking the derivative with respect to $\bm{\Sigma}^{-1}$ in accordance with the standard practice in the unconstrained case \cite[\S 4.2.2 ]{bibby1979multivariate}.
Differentiating with respect to $\bm{\mu}$ and $\bm{\Sigma}^{-1}$ we have:

\begin{align}
\frac{\partial L(\bm{X};\bm{\mu},\bm{\Sigma}) }{\partial \bm{\mu}} &= n\bm{\Sigma}^{-1}(\bar{\bm{x}} - \bm{\mu}) - \bm{\Sigma}^{-1}\bm{\alpha}_2 \nonumber\\
 \frac{\partial L(\bm{X};\bm{\mu},\bm{\Sigma}) }{\partial \bm{\Sigma}^{-1}} &= \frac{1}{2} \left[(n+2\alpha_1)\bm{\Sigma} - \bm{A}(\bm{\mu}) - 2\bm{\alpha}_2\bm{\mu}^{\top}\right]
\end{align}
 and obtain the MLEs as the solution of the following equations:

\begin{align}
\bm{\mu} = \bar{\bm{x}} - \frac{1}{n}\bm{\alpha}_2 ,\quad & \quad \bm{\Sigma} = \frac{\bm{A}(\bm{\mu}) + 2\bm{\alpha}_2\bm{\mu}^{\top}}{n+2\alpha_1} \nonumber\\
\mid\bm{\Sigma}^{-1}\mid = 1 ,\quad & \quad \bm{\Sigma}^{-1}\bm{\mu} = \bm{b}\nonumber
\end{align}

\item  The Lagrangian

\begin{align}
L(\bm{X};\bm{\mu},\bm{\Sigma}) &=c + \frac{n}{2}\log\mid\bm{\Sigma}^{-1}\mid - \frac{1}{2}Tr\left[\bm{A}(\bm{\mu})\bm{\Sigma}^{-1}\right] + \alpha_1(\mid\bm{\Sigma}^{-1}\mid-1) - \bm{\alpha}^{\top}_2\left(\bm{I}_p - \bm{\Sigma}^{-1}\right)\bm{\mu}
\end{align}
and its derivatives with respect to $\bm{\mu}$ and $\bm{\Sigma}^{-1}$ are:

\begin{align}
\frac{\partial L(\bm{X};\bm{\mu},\bm{\Sigma}) }{\partial \bm{\mu}} &= n\bm{\Sigma}^{-1}(\bar{\bm{x}} - \bm{\mu}) - (\bm{I}_p - \bm{\Sigma}^{-1})\bm{\alpha}_2 \nonumber\\
 \frac{\partial L(\bm{X};\bm{\mu},\bm{\Sigma}) }{\partial \bm{\Sigma}^{-1}} &= \frac{1}{2} \left[(n+2\alpha_1)\bm{\Sigma} - \bm{A}(\bm{\mu}) - 2\bm{\alpha}_2\bm{\mu}^{\top}\right]
\end{align}
and obtain the MLE as the solution of the following equations:
\begin{align}
\bm{\mu} = \bar{\bm{x}} - \frac{1}{n}(\bm{I}_p - \bm{\Sigma})\bm{\alpha}_2 ,\quad & \quad \bm{\Sigma} = \frac{\bm{A}(\bm{\mu}) + 2\bm{\alpha}_2\bm{\mu}^{\top}}{n+2\alpha_1} \nonumber\\
\mid\bm{\Sigma}^{-1}\mid = 1 ,\quad & \quad \bm{\Sigma}^{-1}\bm{\mu} = \bm{\mu}\nonumber
\end{align}

\end{enumerate}

\item \textbf{Proof of Lemma \ref{PerMatLem}}
\label{PerMatLemPr}
\begin{enumerate}[label = (\alph*)]
\item Follows from  simple algebra and the definition of eigenvalue:
 \begin{align*}
 (\bm{a}\bm{b}^{\top}+\bm{b}\bm{a}^{\top}) \left(\frac{\bm{a}}{\Vert \bm{a}\Vert} + \frac{\bm{b}}{\Vert \bm{b}\Vert}\right) &=   \left(\bm{a}^{\top}\bm{b} + \Vert \bm{a}\Vert \Vert \bm{b}\Vert\right) \left(\frac{\bm{a}}{\Vert \bm{a}\Vert} + \frac{\bm{b}}{\Vert \bm{b}\Vert}\right)\\
 (\bm{a}\bm{b}^{\top}+\bm{b}\bm{a}^{\top}) \left(\frac{\bm{a}}{\Vert \bm{a}\Vert} - \frac{\bm{b}}{\Vert \bm{b}\Vert}\right) &=   \left(\bm{a}^{\top}\bm{b} - \Vert \bm{a}\Vert \Vert \bm{b}\Vert\right) \left(\frac{\bm{a}}{\Vert \bm{a}\Vert} - \frac{b}{\Vert \bm{b}\Vert}\right)
 \end{align*}

 \item Let $\bm{B} = \bm{B}(\bm{a},\bm{b}) = \bm{a}\bm{b}^{\top} + \bm{b}\bm{a}^{\top}$, and $\lambda_j(\bm{M})$ be the j-th largest eigenvalue of $\bm{M}$ with $\bm{M} = \bm{A}+\bm{B} $. We apply Weyl's Inequality \citep[Theorem 8.2]{bhatia2007perturbation} to obtain
\begin{align*}
\lambda_j(\bm{M}) = \lambda_j(\bm{B} + \bm{A}) &\geq \lambda_j(\bm{B}) + \lambda_p(\bm{A})\\
&\geq \lambda_p(\bm{B}) + \lambda_p(\bm{A}) \\
&= (\bm{a}^{\top}\bm{b} - \Vert \bm{a}\Vert \Vert \bm{b}\Vert ) + \lambda_p(\bm{A}), \quad \text{ (By applying the first part.) }
\end{align*}
 where $\lambda_p(\bm{A})> 0$.   Since $\bm{B}$ is a rank two matrix its at most two non-zero eigenvalues are $(\bm{a}^{\top}\bm{b} \pm \Vert \bm{a}\Vert \Vert \bm{b}\Vert)$.   By applying Cauchy-Schwartz inequality it follows that these non-zero eigenvalues belong to the range $\left(\bm{a}^{\top}\bm{b} + \Vert \bm{a}\Vert \Vert \bm{b}\Vert\right) \in \left[0,2\Vert \bm{a}\Vert \Vert \bm{b}\Vert\right]$ and $\left(\bm{a}^{\top}\bm{b} - \Vert \bm{a}\Vert \Vert \bm{b}\Vert\right) \in \left[-2\Vert \bm{a}\Vert \Vert \bm{b}\Vert,0\right] $. This tells us that $\lambda_1(\bm{B})\geq 0$ and $\lambda_j(\bm{B}) = 0$ for $j=2,3,\dots,p-1$. Weyl's inequality \citep[Theorem 8.2]{bhatia2007perturbation} for $j=2,3,\dots,p-1$,  gives us $$\lambda_j(\bm{M}) \geq \lambda_j(\bm{B}) + \lambda_p(\bm{A}) = \lambda_{p}(\bm{A})>0$$ and  $\lambda_1(\bm{M}) > 0 $ trivially. \par

 This cannot be said for the lowest eigenvalue of $\bm{M}$ i.e. for $j=p$, we cannot say whether $(\bm{a}^{\top}\bm{b} - \Vert \bm{a}\Vert \Vert \bm{b}\Vert ) + \lambda_p(\bm{A})$ is positive or not. It depends on $\lambda_p(\bm{A})$.   Therefore except for the smallest eigenvalue all other eigenvalues of $\bm{M}$ are positive completing the proof.
 \end{enumerate}

\item \label{Condver} \textbf{Verification of Conditions} $\mathcal{F}1-\mathcal{F}4$ and $\mathcal{H}1-\mathcal{H}3$ \citep{aitchison1958maximum}

 Checking the conditions amounts to calculation of second derivative matrix of the likelihood function, which in turn verifies the existence of the third derivative as one of the conditions. Here we present the details of these calculations.

{\bf First Derivative:}

\begin{align}
\frac{\partial l}{\partial \bm{\mu} } = n\bm{\Sigma}^{-1}(\bar{\bm{x}} - \bm{\mu}),  \quad & \quad
 \frac{\partial l}{\partial \bm{\Sigma} } = -\frac{1}{2} \left(n\bm{\Sigma}^{-1} - \bm{\Sigma}^{-1}\bm{A}(\bm{\mu})\bm{\Sigma}^{-1}\right)
\end{align}

{\bf Second Derivative:} The Hessian matrix of the likelihood is:

$$H_{l} = \begin{pmatrix}
\frac{\partial^2 l}{\partial \bm{\mu}^{2}}\biggr\rvert_{p\times p} & \frac{\partial^2 l}{\partial \bm{\mu}\partial \bm{\Sigma}}\biggr\rvert_{p\times p^2}\\
\frac{\partial^2 l}{\partial \bm{\Sigma}\partial \bm{\mu}}\biggr\rvert_{p^2\times p} & \frac{\partial^2 l}{\partial \bm{\Sigma}^2}\biggr\rvert_{p^2\times p^2}
\end{pmatrix}_{p(p+1)\times p(p+1)}. $$ Next, we calculate the four submatrices.

\begin{enumerate}
\item  The first submatrix is \begin{align}
 \frac{\partial^2 l}{\partial \bm{\mu}^{2}} &= -n\bm{\Sigma}^{-1}
 \end{align}

 \item Since,   $\frac{\partial \bm{\Sigma}^{-1}}{\partial \bm{\Sigma}} = - \bm{\Sigma}^{-1}\otimes \bm{\Sigma}^{-1} \text{ and } \frac{\partial \bm{\Sigma}^{-1}(\bar{\bm{x}} - \bm{\mu})}{\partial \bm{\Sigma}^{-1}}  = \bm{I} \otimes (\bm{\bar{x}}-\bm{\mu})$ we obtain the following:
 \begin{align}
 \frac{\partial^2 l}{\partial \bm{\Sigma}\partial \bm{\mu}}\biggr\rvert_{p^2\times p}  &= -\left[\bm{\Sigma}^{-1}\otimes\bm{\Sigma}^{-1}\right]\left[ \bm{I} \otimes (\bm{\bar{x}}-\bm{\mu})\right]
 \end{align}

 \item The third submatrix is \begin{align}
\frac{\partial^2 l}{\partial \bm{\mu}\partial \bm{\Sigma}}\biggr\rvert_{p\times p^2} &= -\frac{1}{2} \frac{\partial }{\partial \bm{\mu}} \text{vec}\left[n\bm{\Sigma}^{-1} - \bm{\Sigma}^{-1}\bm{A}(\bm{\mu})\bm{\Sigma}^{-1}\right]  \nonumber\\
&= \frac{1}{2} \frac{\partial }{\partial \bm{\mu}}  \text{vec}\left[\bm{\Sigma}^{-1}\bm{A}(\bm{\mu})\bm{\Sigma}^{-1}\right]\nonumber\\
&= \frac{1}{2}   \frac{\partial \text{vec}\left(\bm{\Sigma}^{-1}\left[-2n\bar{\bm{x}}\bm{\mu}^{\top} + n\bm{\mu}\bm{\mu}^{\top}\right]\bm{\Sigma}^{-1}\right)}{\partial \bm{\mu}} \nonumber\\
&= \frac{1}{2}\left[-2n\frac{\partial \text{vec}\left[(\bm{\Sigma}^{-1}\bar{\bm{x}})(\bm{\Sigma}^{-1}\bm{\mu})^{\top}\right] }{\partial\bm{\mu}} + n \frac{\partial \text{vec}\left[(\bm{\Sigma}^{-1}\bm{\mu})(\bm{\Sigma}^{-1}\bm{\mu})^{\top}\right] }{\partial\bm{\mu}}\right]\nonumber\\
&=  \frac{1}{2}\left[-2n\frac{\partial (\bm{\Sigma}^{-1}\bm{\mu})^{\top}\otimes (\bm{\Sigma}^{-1}\bar{\bm{x}})^{\top} }{\partial\bm{\mu}} + n \frac{\partial (\bm{\Sigma}^{-1}\bm{\mu})^{\top}\otimes(\bm{\Sigma}^{-1}\bm{\mu})^{\top} }{\partial\bm{\mu}}\right]\nonumber\\
&=  \frac{1}{2}\left[-2n \bm{\Sigma}^{-1}\otimes (\bm{\Sigma}^{-1}\bar{\bm{x}})^{\top} + n \bm{\Sigma}^{-1}\otimes (\bm{\Sigma}^{-1}\bm{\mu})^{\top} + n (\bm{\Sigma}^{-1}\bm{\mu})^{\top} \otimes \bm{\Sigma}^{-1} \right]
\end{align}

\item \noindent Following the calculations of \cite{chaudhuri2007estimation}, we obtain

\begin{align}
\frac{\partial^2 l}{\partial \bm{\Sigma}^2}\biggr\rvert_{p^2\times p^2} &= \frac{1}{2}\left[n\bm{\Sigma}^{-1}\otimes \bm{\Sigma}^{-1} - \left(\bm{\Sigma}^{-1}\bm{A}(\bm{\mu})\bm{\Sigma}^{-1}\right)\otimes \bm{\Sigma}^{-1} - \bm{\Sigma}^{-1}\otimes\left(\bm{\Sigma}^{-1}\bm{A}(\bm{\mu})\bm{\Sigma}^{-1}\right)\right]
\end{align}
\end{enumerate}

 We have shown the existence of the second derivative and  from the above quantities it is evident the the third derivative exists too. Since multivariate normal has all the moments \citep{chacon2015efficient}, so we have essentially verified conditions $\mathcal{F}1-\mathcal{F}4$.
  Next, we verify $\mathcal{H}1-\mathcal{H}3$ for the constraint. The simplest way is to express it as $h(\bm{\mu},\bm{\Sigma}) = \bm{\Sigma}\bm{\mu} - \bm{\mu} = \bm{0}$. Here also we need to check the Hessian matrix of the constraint and its corresponding bound. We can establish the coordinate wise bound to be 1. The details are as follows:

{\bf First Derivative}
\begin{align}
\label{h1thet}
\frac{\partial h}{\partial \bm{\mu}}\biggr\rvert_{p\times p} = \bm{\Sigma} - \bm{I},\quad & \quad \frac{\partial h}{\partial \bm{\Sigma}}\biggr\rvert_{p^2\times p} = \bm{\mu}\otimes \bm{I}
\end{align}

\noindent We denote the first derivative to be $\left(\bm{H}^{1}_{\bm{\theta}}\right)_{p+p^2 \times p}$ with $rank(\bm{H}^{1}_{\bm{\theta}})=p$

{\bf Second Derivative:}
\begin{align}
\frac{\partial^2 h}{\partial \bm{\mu}^2}\biggr\rvert_{p^2\times p} = \bm{0}, \quad & \quad \frac{\partial^2 h}{\partial \bm{\Sigma}\partial\bm{\mu}}\biggr\rvert_{p^2\times p^2} = \bm{I}_{p^2}\nonumber\\
\frac{\partial^2 h}{\partial \bm{\mu}\partial\bm{\Sigma}}\biggr\rvert_{p^3\times p} = \bm{I}\otimes \text{vec}(\bm{I}_p),\quad & \quad \frac{\partial^2 h}{\partial \bm{\Sigma}^2}\biggr\rvert_{p^3\times p^2} = \bm{0}
\end{align}
 This gives us $$\bm{H}^2_{\bm{\theta}}\biggr\rvert_{(p^2+p^3)\times (p+p^2)} = \left(\left(\frac{\partial^2 h}{\partial \theta_i \partial \theta_j}\right)\right)$$ with $rank(\bm{H}^2_{\bm{\theta}})=p+p^2$

\item \textbf{Proof of Lemma \ref{DirDerLem}:}
\label{PrDirDerLem}
We assume that $\bm{\mu}$ is fixed and are interested in calculating the directional derivative of the Lagrangian in (\ref{LagM}) as a function of $\bm{\Sigma}$ only in the direction of a symmetric matrix $\bm{D}$. Let us set $L(\bm{\mu},\bm{\Sigma}\mid\bm{X}) = f(\bm{\Sigma})$. By the definition of directional derivative,

 $$\nabla_{\bm{D}} f = \lim_{h\to 0}\frac{f(\bm{\Sigma}+h\bm{D})-f(\bm{\Sigma})}{h}$$ where $h$ is a scalar. Now since the terms in the Lagrangian are additive, we calculate the directional derivative of each term separately.

 \begin{enumerate}
\item First we focus on the first term ignoring the constant $f^{1}=\log\mid\bm{\Sigma}\mid$.

 \begin{align*}
\nabla_{\bm{D}} f^{1} &= \lim_{h\to 0} \frac{1}{h} \log \left[\frac{\mid\bm{\Sigma}+h\bm{D}\mid}{\mid\bm{\Sigma}\mid}\right] = \lim_{h\to 0} \frac{1}{h} \log\left[\frac{\mid\bm{\Sigma}\mid\mid\bm{I}+h\bm{D}\bm{\Sigma}^{-1}\mid}{\mid\bm{\Sigma}\mid}\right] \\
&= \lim_{h\to 0} \frac{1}{h} \log\left[\mid\bm{I}+h\bm{D}\bm{\Sigma}^{-1}\mid\right]\\
&= \lim_{h\to 0} \frac{1}{h} \sum_{i=1}^{p }\log\left[1+ h\lambda_i(\bm{D}\bm{\Sigma}^{-1})\right] \quad\text{(Here $\lambda_i(\bm{D})$ denote the ith eigenvalue of $\bm{D}$) }\\
&= \lim_{h\to 0} \frac{1}{h} \left[\sum_{i=1}^p h\lambda_i(\bm{D}\bm{\Sigma}^{-1}) + \sum_{i=1}^{p } \left\{\log\left[1+ h\lambda_i(\bm{D}\bm{\Sigma}^{-1})\right] - h\lambda_i(\bm{D}\bm{\Sigma}^{-1})\right\}\right]= Tr\left[\bm{D}\bm{\Sigma}^{-1}\right]
 \end{align*}

\item The second term ignoring the constant $f^2 = Tr\left[\bm{S}\bm{\Sigma}^{-1}\right]$.

\begin{align*}
\nabla_{\bm{D}} f^2 &= \lim_{h\to 0} \frac{1}{h} Tr\left[\bm{S}\left\{(\bm{\Sigma} + h{\bm{D}})^{-1} - \bm{\Sigma}^{-1}\right\}\right]\\
&=\lim_{h\to 0} \frac{1}{h} Tr\left[\bm{S} \left\{\bm{\Sigma}^{-1} - \bm{\Sigma}^{-1}(h^{-1}\bm{D}^{-1} - \bm{\Sigma}^{-1})^{-1}\bm{\Sigma}^{-1} - \bm{\Sigma}^{-1}\right\} \right]\\
&=\lim_{h\to 0} \frac{1}{h} Tr\left[\bm{S} \left\{ - \bm{\Sigma}^{-1}(h^{-1}\bm{D}^{-1} + \bm{\Sigma}^{-1})^{-1}\bm{\Sigma}^{-1} \right\} \right]\\
&= \lim_{h\to 0} \frac{1}{h} Tr\left[\bm{S} \left\{ - h\bm{\Sigma}^{-1}(\bm{D}^{-1} + h\bm{\Sigma}^{-1})^{-1}\bm{\Sigma}^{-1} \right\} \right]\\
&= \lim_{h\to 0} Tr\left[-\bm{S}\bm{\Sigma}^{-1}(\bm{D}^{-1}+h\bm{\Sigma}^{-1})^{-1}\bm{\Sigma}^{-1}\right]\\
&= Tr\left[-\bm{S}\bm{\Sigma}^{-1}\bm{D}\bm{\Sigma}^{-1}\right]\\
\end{align*}

\item The third term without the constant is $f^3 = Tr\left[(\bar{\bm{x}}-\bm{\mu})(\bar{\bm{x}}-\bm{\mu})^{\top}\bm{\Sigma}^{-1}\right] $. The same calculation shows that
$\nabla_{\bm{D}} f^3 = Tr\left[-(\bar{\bm{x}}-\bm{\mu})(\bar{\bm{x}}-\bm{\mu})^{\top}\bm{\Sigma}^{-1}\bm{D}\bm{\Sigma}^{-1}\right] $

\item The fourth term $f^4 = \bm{\alpha}_2^{\top}(\bm{\Sigma}\bm{\mu} - \bm{\mu})$

\begin{align*}
\nabla_{\bm{D}} f^4 &= \lim_{h\to 0} \frac{\left\{\bm{\alpha}_2^{\top}\left(\bm{\Sigma}+h\bm{D}\right)\bm{\mu} - \bm{\alpha}_2^{\top}\bm{\mu} - \bm{\alpha}_2^{\top}\bm{\Sigma}\bm{\mu} + \bm{\alpha}_2^{\top}\bm{\mu}\right\}}{h} = \lim_{h\to 0} \frac{\left\{\bm{\alpha}_2^{\top}\left(\bm{\Sigma}+h\bm{D}\right)\bm{\mu}  - \bm{\alpha}_2^{\top}\bm{\Sigma}\bm{\mu} \right\}}{h}\\
&= \lim_{h\to 0} \frac{\bm{\alpha}_2^{\top}\left(\bm{\Sigma}+h\bm{D} - \bm{\Sigma}\right)\bm{\mu} }{h} = \bm{\alpha}_2^{\top}\bm{D}\bm{\mu}
\end{align*}

 \end{enumerate}

With these calculations the final directional derivative of the Lagrangian function is:

 \begin{align}
 \nabla_{\bm{D}} f &= -\frac{n}{2}Tr\left[\bm{D}\bm{\Sigma}^{-1}\right] + \frac{n}{2} Tr\left[\bm{S}\bm{\Sigma}^{-1}\bm{D}\bm{\Sigma}^{-1}\right] + \frac{n}{2}Tr\left[\bm{B}\bm{\Sigma}^{-1}\bm{D}\bm{\Sigma}^{-1}\right]+\bm{\alpha}_2^{\top}\bm{D}\bm{\mu}\nonumber\\
 &=  -\frac{n}{2}Tr\left[\bm{D}\bm{\Sigma}^{-1}\right] + \frac{n}{2} Tr\left[(\bm{S}+\bm{B})\bm{\Sigma}^{-1}\bm{D}\bm{\Sigma}^{-1}\right] +\bm{\alpha}_2^{\top}\bm{D}\bm{\mu}
 \end{align}
where $\bm{B} = (\bar{\bm{x}}-\bm{\mu})(\bar{\bm{x}}-\bm{\mu})^{\top}$

Now, we calculate the second directional derivative in the direction $\bm{C}$ which is also a symmetric matrix. The first derivative denoted by $\nabla_{\bm{D}} f$ has two terms as a function of $\bm{\Sigma}$. The corresponding notation for second directional derivative is $$\nabla_{\bm{C}}\nabla_{\bm{D}}f = \lim_{h\to 0} \frac{\nabla_{\bm{D}} f(\bm{\Sigma}+h\bm{C}) - \nabla_{\bm{D}} f(\bm{\Sigma})}{h}.$$
We will calculate the directional derivative of each of the two terms.
\begin{enumerate}
\item The first term ignoring the constant is $\nabla_{\bm{D}} f^{1} = Tr\left[\bm{D}\bm{\Sigma}^{-1}\right]$. This is same as the second term of the original likelihood function. So by applying the same formula we obtain: $\nabla_{\bm{C}}\nabla_{\bm{D}} f^{1} = -Tr\left[\bm{D}\bm{\Sigma}^{-1}\bm{C}\bm{\Sigma}^{-1}\right] = -Tr\left[\bm{\Sigma}\bm{\Sigma}^{-1}\bm{D}\bm{\Sigma}^{-1}\bm{C}\bm{\Sigma}^{-1}\right]$

\item The second term ignoring the constant is $\nabla_{\bm{D}} f^{2} = Tr\left[(\bm{S}+\bm{B})\bm{\Sigma}^{-1}\bm{D}\bm{\Sigma}^{-1}\right]$. By Woodbury-Sherman matrix formula:

\begin{align*}
&(\bm{\Sigma}+h\bm{C})^{-1}\bm{D}(\bm{\Sigma}+h\bm{C})^{-1} -  \bm{\Sigma}^{-1}\bm{D}\bm{\Sigma}^{-1}  \\
&=\left[\bm{\Sigma}^{-1} - \bm{\Sigma}^{-1}(h^{-1}\bm{C}^{-1} + \bm{\Sigma}^{-1})^{-1}\bm{\Sigma}^{-1} \right]\bm{D}\left[\bm{\Sigma}^{-1} - \bm{\Sigma}^{-1}(h^{-1}\bm{C}^{-1} + \bm{\Sigma}^{-1})^{-1}\bm{\Sigma}^{-1} \right] - \bm{\Sigma}^{-1}\bm{D}\bm{\Sigma}^{-1}\\
&= \left[\bm{\Sigma}^{-1} - h\bm{\Sigma}^{-1}(\bm{C}^{-1} + h \bm{\Sigma}^{-1})^{-1}\bm{\Sigma}^{-1} \right]\bm{D}\left[\bm{\Sigma}^{-1} - h\bm{\Sigma}^{-1}(\bm{C}^{-1} + h \bm{\Sigma}^{-1})^{-1}\bm{\Sigma}^{-1} \right] - \bm{\Sigma}^{-1}\bm{D}\bm{\Sigma}^{-1}\\
&= - h\bm{\Sigma}^{-1}(\bm{C}^{-1} + h \bm{\Sigma}^{-1})^{-1}\bm{\Sigma}^{-1} \bm{D}\bm{\Sigma}^{-1} - h\bm{\Sigma}^{-1}\bm{D}\bm{\Sigma}^{-1}(\bm{C}^{-1} + h \bm{\Sigma}^{-1})^{-1}\bm{\Sigma}^{-1} + \mathcal{O}(h^2)
\end{align*}

Using this result, we get:

\begin{align*}
\nabla_{\bm{C}} \nabla_{\bm{D}} f^2 &= \lim_{h\to 0} \frac{Tr\left[(\bm{S}+\bm{B})(\bm{\Sigma}+h\bm{C})^{-1}\bm{D}(\bm{\Sigma}+h\bm{C})^{-1}\right]  -  Tr\left[(\bm{S}+\bm{B})\bm{\Sigma}^{-1}\bm{D}\bm{\Sigma}^{-1}\right]}{h}\\
&=  \lim_{h\to 0} \frac{Tr\left[(\bm{S}+\bm{B})\left\{(\bm{\Sigma}+h\bm{C})^{-1}\bm{D}(\bm{\Sigma}+h\bm{C})^{-1} -  \bm{\Sigma}^{-1}\bm{D}\bm{\Sigma}^{-1}\right\}\right]}{h}\\
&=-\lim_{h\to 0}Tr\left[(\bm{S}+\bm{B})\bm{\Sigma}^{-1}(\bm{C}^{-1} + h \bm{\Sigma}^{-1})^{-1}\bm{\Sigma}^{-1} \bm{D}\bm{\Sigma}^{-1}\right] \\
&\qquad\qquad- \lim_{h\to 0} Tr\left[(\bm{S}+\bm{B})\bm{\Sigma}^{-1}\bm{D}\bm{\Sigma}^{-1}(\bm{C}^{-1} + h \bm{\Sigma}^{-1})^{-1}\bm{\Sigma}^{-1}\right]\\
&= -Tr\left[(\bm{S}+\bm{B})\bm{\Sigma}^{-1}\bm{C} \bm{\Sigma}^{-1} \bm{D}\bm{\Sigma}^{-1}\right] - Tr\left[(\bm{S}+\bm{B})\bm{\Sigma}^{-1}\bm{D}\bm{\Sigma}^{-1}\bm{C}\bm{\Sigma}^{-1}\right]\\
&= -Tr\left[(\bm{S}+\bm{B})\bm{\Sigma}^{-1}\bm{C} \bm{\Sigma}^{-1} \bm{D}\bm{\Sigma}^{-1}\right] - Tr\left[(\bm{S}+\bm{B})\bm{\Sigma}^{-1}(\bm{D}\bm{\Sigma}^{-1}\bm{C})^{\top}\bm{\Sigma}^{-1}\right]\\
&= - Tr\left[2(\bm{S}+\bm{B})\bm{\Sigma}^{-1}\bm{C} \bm{\Sigma}^{-1} \bm{D}\bm{\Sigma}^{-1}\right]
\end{align*}

\end{enumerate}

Adding these two we obtain the final directional derivative to be:

\begin{align}
\nabla_{\bm{C}}\nabla_{\bm{D}}f &= \frac{n}{2}Tr\left[\bm{\Sigma}\bm{\Sigma}^{-1}\bm{D}\bm{\Sigma}^{-1}\bm{C}\bm{\Sigma}^{-1}\right] - \frac{n}{2} Tr\left[2(\bm{S}+\bm{B})\bm{\Sigma}^{-1}\bm{C} \bm{\Sigma}^{-1} \bm{D}\bm{\Sigma}^{-1}\right]\nonumber\\
&= -\frac{n}{2} Tr\left[\left\{2(\bm{S}+\bm{B}) - \bm{\Sigma}\right\}\bm{\Sigma}^{-1}\bm{C} \bm{\Sigma}^{-1} \bm{D}\bm{\Sigma}^{-1}\right]
\end{align}

Now if we assume that the mean vector $\bm{\mu}$ is estimated by $\bar{\bm{x}}$, then $\bm{B} = \bm{0}$. If we further assume that the estimate of the covariance matrix lies within the set $D_{2S} = \left\{\bm{\Sigma}\text{ is pd, } 0\prec \bm{\Sigma} \prec 2\bm{S}\right\}$ and $\bm{C} = \bm{D}$, then
\begin{align}
\nabla_{\bm{D}}\nabla_{\bm{D}}f &= -\frac{n}{2}Tr\left[ \bm{\Sigma}^{-1/2}\left\{2\bm{S} - \bm{\Sigma}\right\}\bm{\Sigma}^{-1/2}\bm{\Sigma}^{-1/2}\bm{D}\bm{\Sigma}^{-1/2}\bm{\Sigma}^{-1/2}\bm{D}\bm{\Sigma}^{-1/2}\right] \nonumber\\
&= -\frac{n}{2}Tr\left[\bm{\Sigma}^{-1/2}\bm{D}\bm{\Sigma}^{-1/2}\bm{\Sigma}^{-1/2}\left\{2\bm{S} - \bm{\Sigma}\right\}\bm{\Sigma}^{-1/2}\bm{\Sigma}^{-1/2}\bm{D}\bm{\Sigma}^{-1/2}\right] \leq 0
\end{align}

 Therefore, within $D_{2S}$ the constrained likelihood is strictly concave, but outside this set that is not the case as shown by the following
   counter-example:

       If $\bm{\Sigma}\notin D_{2S}$, then there exists a $\bm{u}$ such that $\bm{u}^{\top}(2\bm{S} - \bm{\Sigma})\bm{u}\leq 0$. Choosing $\bm{D} = \bm{\Sigma}\bm{u}\bm{u}^{\top}\bm{\Sigma}$ then

\begin{align}
\nabla_{\bm{D}}\nabla_{\bm{D}}f &= -\frac{n}{2}Tr\left[\bm{\Sigma}^{1/2}\bm{u}\bm{u}^{\top}(2\bm{S} - \bm{\Sigma})\bm{u}\bm{u}^{\top}\bm{\Sigma}^{1/2}\right]
=-\frac{n}{2} \bm{u}^{\top}\bm{\Sigma}\bm{u} \bm{u}^{\top}(2\bm{S}-\bm{\Sigma})\bm{u} \geq 0,
\end{align}
which completes the proof.

\item \textbf{Covariance error bound in Lemma \ref{AlgoM}:}\label{L2AlgoMEx} The error of a new estimate can be bounded in the following way:
\begin{align}
\label{AlgoSigEr}
\left\Vert\bm{\Sigma}-\widehat{\bm{\Sigma}}^*\right\Vert_{\mathcal{F}} &\leq \left\Vert\bm{\Sigma}-\widehat{\bm{\Sigma}}_{map}\right\Vert_{\mathcal{F}} + \left\Vert\widehat{\bm{\Sigma}}_{map}-\widehat{\bm{\Sigma}}^*\right\Vert_{\mathcal{F}}\nonumber\\
\left\Vert \widehat{\bm{\Sigma}}_{map} - \widehat{\bm{\Sigma}}^*\right\Vert_{\mathcal{F}} &= \left\Vert \left(1-\frac{1}{\lambda_P}\right)\sum_{\substack{i=1\\i\neq i_0}}^d \lambda_i \bm{P}_i\bm{P}_i^{\top} + \left(\frac{\lambda_{i_0}}{c^2_{0i_0}} - 1\right)\widehat{\bm{\mu}}^*\widehat{\bm{\mu}}^{*\top}\right\Vert_{\mathcal{F}}\nonumber\\
\text{ Applying triangle inequality : } &\nonumber\\
&\leq \left\Vert\left(1-\frac{1}{\lambda_P}\right)\sum_{\substack{i=1\\i\neq i_0}}^d \lambda_i \bm{P}_i\bm{P}_i^{\top}\right\Vert_{\mathcal{F}}+\left\Vert \left(\frac{\lambda_{i_0}}{c^2_{0i_0}} - 1\right)\widehat{\bm{\mu}}^*\widehat{\bm{\mu}}^{*\top}\right\Vert_{\mathcal{F}}\nonumber\\
&=\left| \left(1-\frac{1}{\lambda_P}\right)\right|\left\Vert\sum_{\substack{i=1\\i\neq i_0}}^d \lambda_i \bm{P}_i\bm{P}_i^{\top}\right\Vert_{\mathcal{F}}+\left|\left(\frac{\lambda_{i_0}}{c^2_{0i_0}} - 1\right)\right|\left\Vert \widehat{\bm{\mu}}^*\widehat{\bm{\mu}}^{*\top}\right\Vert_{\mathcal{F}}
\end{align}

\item \textbf{Proof of Equation \ref{muEs}:}
This follows from standard regression OLS estimate.

\item \textbf{Proof of Equation \ref{LEs}:}
\label{EVEstPr}
\begin{align}
f\left(\lambda^'_{\mathcal{S}}\right) &= Tr\left[\left(\bm{\Sigma}-\widehat{\bm{\Sigma}}^*\right)^2\right]\nonumber\\
\frac{\partial f\left(\lambda^'_{\mathcal{S}}\right)}{\partial \lambda_{i_k}} &= -2.\text{Tr}\left[\left(\bm{\Sigma}-\widehat{\bm{\Sigma}}^*\right)\left(\bm{b}_k\bm{b}_k^{\top}\right)\right]=0\nonumber\\
\implies \text{Tr}\left[\bm{b}_k^{\top}\left(\bm{\Sigma}-\widehat{\bm{\Sigma}}^*\right)\bm{b}_k\right] &= 0\nonumber\\
\implies \bm{b}_k^{\top}\bm{\Sigma} \bm{b}_k &= \bm{b}_k^{\top}\widehat{\bm{\Sigma}}^* \bm{b}_k \nonumber\\
\implies \widehat{\lambda^'}_{i_k} &= \bm{b}_k^{\top}\bm{\Sigma} \bm{b}_k
\end{align}

\end{enumerate}

\section{ Explicit Calculation of the Lagrange Multiplier}
\label{MLESAiter}

    We consider finding the MLE under constraints  for an exponential family of distributions:

 \begin{align*}
 f(\bm{X};\bm{\theta}) &= \exp\left[ b_0(\bm{X}) + \sum_{i=1}^q\theta_iT_i(\bm{X}) - a(\bm{\theta}) \right]
 \end{align*}
 where $\bm{\theta} = (\bm{\theta}_1,\dots,\bm{\theta}_q)$ is the vector of natural parameters and $\bm{T}(\bm{X}) = (T_1(\bm{X}),T_2(\bm{X}),\dots,T_q(\bm{X}))$ is their complete and sufficient statistics with the following \citep{lehmann2006theory}
  $$\mathbb{E}\left[T(\bm{X})\right] = \frac{\partial a(\bm{\theta})}{\partial \bm{\theta}} = \bm{m} \quad\text{and } \quad\text{Cov}\left[T(\bm{X})\right] = \frac{\partial^2 a(\bm{\theta})}{\partial\bm{\theta}\partial\bm{\theta}^{\top}} = \bm{V}.$$

  Let the constraint on the parameters be expressed as a function $h(\bm{m}) = 0$ where $h(\bm{m}):\mathbb{R}^q \to \mathbb{R}$ and  both $\bm{m}$ and $\bm{V}$ are functions of $\bm{\theta}$. Differentiating the Lagrangian function
  \begin{align*}
  w(\bm{X};\bm{\theta},\alpha_2) &= \log f(\bm{X};\bm{\theta}) + \alpha_2 h(\bm{m}) \propto \bm{\theta}^{\top}T(\bm{X}) - a(\bm{\theta})+ \alpha_2 h(\bm{m})
  \end{align*}

 \noindent with respect to $\bm{\theta}$ and equating it to zero, we obtain
  \begin{align}
  \frac{\partial w(\bm{X};\bm{\theta},\alpha_2)}{\partial \bm{\theta}} & = T(\bm{X}) - \frac{\partial a(\bm{\theta})}{\partial \bm{\theta}} + \alpha_2 \frac{\partial h(\bm{m})}{\partial \bm{\theta}} = 0\nonumber\\
  \label{AMTS}
  \bm{m} &= T(\bm{X}) + \alpha_2 \nabla \bm{m}(\bm{\theta}) \nabla h(\bm{m})
  \end{align}

\noindent where $ \nabla \bm{m}(\bm{\theta})$ is a $q\times q$ gradient matrix and $\nabla h(\bm{m})$ denotes a $q\times 1$ gradient vector. The algorithms proposed in Section \ref{PMC} approximates the estimate of Lagrange multiplier within the iterations  so that the iterations are free of the nuisance
Lagrange parameters, see \cite{matthews1995maximum}, \cite{strydom2012maximum}. The Taylor series expansion of $h(\bm{m})$ around $T(\bm{X})$ and the approximation of unknown $\gamma$ is performed as follows:

\begin{align*}
0 = h(\bm{m}) &= h(T) + \alpha_2 \nabla h(\bm{m})^{\top} \nabla \bm{m}(\bm{\theta}) \nabla h(T) + o(\vert\vert \bm{m} - T\vert\vert) \\
\alpha_2 &= - \left[\nabla h(\bm{m})^{\top} \nabla \bm{m}(\bm{\theta}) \nabla h(T)\right]^{-1} h(T)
\end{align*}
 Substituting this value in  (\ref{AMTS}) we obtain the final approximations for $\bm{m}$ to be the equation (\ref{SANRI}).

\noindent \textbf{Example:} We focus on the constraint $\bm{\Sigma}\bm{\mu} = \bm{\mu}$ under normal distribution as an example of the  general set up described above. We can rewrite the constraint  in terms of a suitable differentiable $h$, and as a function of the expectation of the sufficient statistic.
The log-likelihood of normal distribution as in (\ref{loglik}), can also be expressed in terms of natural parameters in the following way:
\begin{align*}
l(\bm{\mu},\bm{\Sigma}\mid \bm{X}) &\propto n\bm{\mu}^{\top}\bm{\Sigma}^{-1}\bar{\bm{x}} - \frac{n}{2}\text{Tr}\left[\bm{\Sigma}^{-1}\left(\frac{1}{n}\sum_{i=1}^p \bm{x}_i\bm{x}_i^{\top}\right)\right] - \frac{n}{2}\bm{\mu}^{\top}\bm{\Sigma}^{-1}\bm{\mu} - \frac{n}{2}\log[\text{det}(2\pi\bm{\Sigma})] \\
&= \bm{\theta}^{\top}T - a(\bm{\theta})
\end{align*}
 Here $T$, the sufficient statistic and $\bm{\theta}$, the natural parameter are:
\begin{align*}
T(\bm{X}) = \begin{pmatrix}
\bar{\bm{x}}\\
\text{vec}\left(\frac{1}{n}\sum_{i=1}^p \bm{x}_i\bm{x}_i^{\top}\right)
\end{pmatrix} ,\qquad &\qquad \bm{\theta} = \begin{pmatrix}
n\bm{\Sigma}^{-1}\bm{\mu}\\
-\frac{n}{2}\text{vec}(\bm{\Sigma}^{-1})
\end{pmatrix}
\end{align*}
with
\begin{align*}
\mathbb{E}(T) = \bm{m}= \begin{pmatrix}
\bm{\mu}\\
\text{vec}(\bm{\Sigma}+\bm{\mu}\bm{\mu}^{\top})
\end{pmatrix} = \begin{pmatrix}
\bm{m}_1\\
\bm{m}_2
\end{pmatrix}  ,\quad
 \text{cov}(T) = V=\begin{pmatrix}
\bm{V}_{11} & \bm{V}_{12}\\
\bm{V}_{21} & \bm{V}_{22}
\end{pmatrix}
\end{align*}
\noindent where
\begin{align*}
\bm{V}_{11} = \frac{1}{n}\bm{\Sigma},\quad &\quad
\bm{V}_{12} = \frac{1}{n}\left(\bm{\Sigma}\otimes\bm{\mu}+\bm{\mu}\otimes\bm{\Sigma}\right)\\
\bm{V}_{21} = V_{21}^{\top},\quad & \quad
\bm{V}_{22} = \frac{1}{n}\left(\bm{I}_{p^2} + \bm{K}\right)\left[\bm{\Sigma}\otimes\bm{\Sigma} + \bm{\Sigma}\otimes\bm{\mu}\bm{\mu}^{\top}+\bm{\mu}\bm{\mu}^{\top}\otimes\bm{\Sigma}\right]
\end{align*}

The matrix $\bm{K}$ is given by $\bm{K}=\sum_{i,j=1}^p \bm{H}_{ij}\otimes\bm{H}_{ij}$,  where $\bm{H}_{ij}:$ zero matrix except $(i,j)$-th element, $h_{ij}=1$.\par

\textbf{Proof of the Form of $h$ in (\ref{g}):} The condition  $Tr[(\bm{\Sigma}-\bm{I}_p)\bm{R}_{\mu}]=\sum_{i=1}^p(\bm{\Sigma}-\bm{I}_p)_{i.}\bm{\mu}=0$, where $\bm{R}_{\mu} = \bm{\mu}\otimes \mathbb{1}^{\top}$ and $\mathbb{1} = [1,1,\dots,1]^{\top}$ will be written in the form  $h(\bm{m}) = Tr[(\bm{\Sigma}-\bm{I}_p)\bm{R}_{\mu}]$. We know that $\text{vec}(\bm{R}_{\mu})=\mathbb{1}\otimes\bm{\mu}$.

\begin{align*}
\bm{m}_2 &=\text{vec}(\bm{\Sigma}+\bm{\mu}\bm{\mu}^{\top})= \text{vec}(\bm{\Sigma})+\text{vec}(\bm{\mu}\bm{\mu}^{\top})
= \text{vec}(\bm{\Sigma}) + \bm{m}_1\otimes \bm{m}_1\\
h(\bm{m}) &= \left[\bm{m}_2 - \bm{m}_1\otimes\bm{m}_1 - \text{vec}(\bm{I}_p)\right]^{\top}\left(\mathbb{1}\otimes\bm{m}_1\right)
=\left[\text{vec}(\bm{\Sigma}) - \text{vec}(\bm{I}_p)\right]^{\top}\left(\mathbb{1}\otimes\bm{\mu}\right)\\
&=\text{vec}\left(\bm{\Sigma}-\bm{I}_p\right)^{\top}\text{vec}(\bm{R}_{\mu})=Tr\left[\left(\bm{\Sigma}-\bm{I}_p\right)\bm{R}_{\mu}\right]
\end{align*}

     The iteration in  (\ref{SANRI}) requires $\nabla h$ and $\nabla m$. Note that $\nabla m = V$  and $\nabla h$ is calculated as follows:

\begin{align*}
\nabla h &= \left[\frac{\partial h}{\partial \bm{m}}\right]_{(p^2+p) \times 1} = \begin{pmatrix}
\frac{\partial h}{\partial \bm{m}_1}\\
\frac{\partial h}{\partial \bm{m}_2}
\end{pmatrix}\\\\
\shortintertext{where}\\
\frac{\partial h}{\partial \bm{m}_1} &= \frac{\partial \left(\bm{m}_2 - \text{vec}(\bm{I}_p)\right)^{\top}(\mathbb{1}\otimes\bm{m}_1)}{\partial \bm{m}_1} - \frac{(\bm{m}_1\otimes \bm{m}_1)^{\top}(\mathbb{1}\otimes\bm{m}_1)}{\partial \bm{m}_1}\\
&= \left(\mathbb{1}\otimes \bm{I}_p\right)^{\top}\left(\bm{m}_2 -\text{vec}(\bm{I}_p)\right) - \left(\mathbb{1}\otimes \bm{I}_p\right)^{\top} \left(\bm{m}_1\otimes\bm{m}_1\right) \\
& \qquad\qquad\qquad - \left(\bm{m}_1\otimes\bm{I}_p+\bm{I}_p\otimes \bm{m}_1\right)^{\top} \left( \mathbb{1}\otimes \bm{m}_1\right)\\
\frac{\partial h}{\partial \bm{m}_2} &= \mathbb{1}\otimes \bm{m}_1.
\end{align*}

{\bf Algorithm 2:}
The detailed steps for finding the constrained mle in an exponential family is:

\begin{itemize}
\item[Step 1.] Start with an initial value $T_0$, the vector of observed canonical statistics.
\item[Step 2.] Set $T=T_0$
\item[Step 3A.] For the $l-$th iteration of $\bm{m}$:
$\bm{m}^{(l)}=T$ and calculate $\nabla h\left(\bm{m}^{(l)}\right)$ and $\bm{V}^{(l)}$ as a function of $\bm{m}^{(l)}$ and $T$.

\item[Step 3B.] For the $k-$th iteration of $T$:
\begin{enumerate}
\item  calculate $h\left(T^{(k)}\right)$, $\nabla h\left(T^{(k)}\right)$.
\item use (\ref{SANRI}) to update:
 $$T^{(k+1)} = T^{(k)} - \bm{V}^{(l)}\nabla h(\bm{m}^{(l)})\frac{h\left(T^{(k)}\right)}{[\nabla h\left(\bm{m}^{(l)}\right)^{\top}\bm{V}^{(l)}\nabla h\left(T^{(k)}\right)]}.$$

\item If $\left\Vert T^{(k+1)} - T^{(k)}\right\Vert\leq \epsilon$ for some fixed number $\epsilon$ determining accuracy, set
 $T=T^{(k+1)}$  break the loop and go to Step 3A, else repeat the loop.
\end{enumerate}
\item[Step 4.] If  $\left\Vert T - \bm{m}^{(l)}\right\Vert\leq \epsilon$ the convergence is attained and the constrained MLE estimate is $\bm{m} = \bm{m}^{(l)}$.
\end{itemize}

\end{document}